\documentclass[%
 reprint,
superscriptaddress,
nofootinbib,
amsmath,amssymb,
aps,
pra,
]{revtex4-1}
\usepackage[dvipsnames]{xcolor}
\usepackage[english]{babel}
\usepackage{cancel}
\usepackage{soul}
\usepackage{upgreek,xspace}
 \usepackage[dvipsnames]{xcolor}
\usepackage[english]{babel}
\usepackage[shortlabels]{enumitem}
 
\makeatletter
\def\bbl@set@language#1{%
  \edef\languagename{%
    \ifnum\escapechar=\expandafter`\string#1\@empty
    \else\string#1\@empty\fi}%
  \@ifundefined{babel@language@alias@\languagename}{}{%
    \edef\languagename{\@nameuse{babel@language@alias@\languagename}}%
  }%
  \select@language{\languagename}%
  \expandafter\ifx\csname date\languagename\endcsname\relax\else
    \if@filesw
      \protected@write\@auxout{}{\string\select@language{\languagename}}%
      \bbl@for\bbl@tempa\BabelContentsFiles{%
        \addtocontents{\bbl@tempa}{\xstring\select@language{\languagename}}}%
      \bbl@usehooks{write}{}%
    \fi
  \fi}
\newcommand{\DeclareLanguageAlias}[2]{%
  \global\@namedef{babel@language@alias@#1}{#2}%
}
\makeatother

\DeclareLanguageAlias{en}{english}

\usepackage{graphicx}
\usepackage{amsmath}
\usepackage{array}
\usepackage{amssymb}
\usepackage{dcolumn}
\usepackage{bm}
\let\savecorresponds\corresponds
\let\corresponds\relax
\usepackage{mathabx}
\usepackage{enumitem} 
\usepackage{tabularx}
\usepackage{soul}
\let\corresponds\savecorresponds

\renewcommand{\vec}[1]{\mathbf{#1}}

\def\ket#1{ | #1 \rangle}
\def\bra#1{{\langle #1 |  }}
\def\vec#1{\boldsymbol{#1}}

\def\pd2v#1#2#3{\frac{\partial^2 #1}{\partial #2 \partial #3}}
\def\abs#1{\left| #1 \right|}

\def \vec#1{\mathbf{#1}}
\def \2x2mat#1#2#3#4{
\left( \begin{array}{cc}
#1 &  #2 \\  #3 &  #4
\end{array} \right)
}

\begin{document}

\preprint{APS/123-QED}

\title{Polarisation entanglement-enabled quantum holography}

\author{Hugo Defienne}
\email[Corresponding authors: ]{hugo.defienne@glasgow.ac.uk ; daniele.faccio@glasgow.ac.uk.}
\affiliation{School of Physics and Astronomy, University of Glasgow, Glasgow G12 8QQ, UK\
}%
\author{Bienvenu Ndagano}%
\affiliation{School of Physics and Astronomy, University of Glasgow, Glasgow G12 8QQ, UK\
}%
\author{Ashley Lyons}%
\affiliation{School of Physics and Astronomy, University of Glasgow, Glasgow G12 8QQ, UK\
}%
\author{Daniele Faccio$^{1,*}$}%

\date{\today}

\begin{abstract}
Holography is a cornerstone characterisation and imaging technique that can be applied to the full electromagnetic spectrum, from X-rays to radio waves or even particles such as neutrons. The key property in all these holographic approaches is coherence that is required to extract the phase information through interference with a reference beam – without this, holography is not possible. Here we introduce a holographic imaging approach that operates on intrinsically incoherent and unpolarised beams, so that no phase information can be extracted from a classical interference measurement. Instead, the holographic information is encoded in the second order coherence of entangled states of light. Using spatial-polarisation hyper-entangled photons pairs, we remotely reconstruct phase images of complex objects. Information is encoded into the polarisation degree of the entangled state, allowing us to image through dynamic phase disorder and even in the presence of strong classical noise, with enhanced spatial resolution compared to classical coherent holographic systems. Beyond imaging, quantum holography quantifies hyper-entanglement distributed over $10^4$ modes via a spatially-resolved Clauser-Horne-Shimony-Holt inequality measurement, with applications in quantum state characterisation.
\end{abstract}

\maketitle

Holography is an essential tool of modern optics~\cite{gabor_new_1948}, at the origin of many applications for microscopic imaging~\cite{marquet_digital_2005}, optical security~\cite{refregier_optical_1995} and data storage~\cite{heanue_volume_1994}. In this respect, holographic interferometry is a widely-used technique that exploits optical interference to retrieve the phase component of a classical optical field through intensity measurements. For example, phase-shifting holography~\cite{yamaguchi_phase-shifting_1997} uses four intensity images $I_\theta$ ($\theta \in \{0,\pi/2,\pi,3 \pi/2 \}$) of a reference optical field $a e^{i \theta}$ interfering with an unknown field $b e^{i \phi}$ to reconstruct the phase profile
\begin{equation}
\label{equ1}
\phi = \arg \left[ I_0-I_\pi + i (I_{\pi/2}-I_{3\pi/2}) \right].
\end{equation} 
Maintaining optical coherence between interfering fields is therefore essential in all holographic protocols. Mechanical instabilities, random phase disorder and the presence of stray light are examples of phenomena that degrade light coherence and hinder the phase reconstruction process. 

Whilst holography is based on classical interference of light waves, the quantum properties of light have inspired a range of new imaging modalities~\cite{moreau_imaging_2019-1} including interaction-free~\cite{white_``interaction-free_1998,pittman_optical_1995} and induced-coherence imaging~\cite{lemos_quantum_2014}, as well as  sensitivity-enhanced~\cite{nasr_demonstration_2003,brida_experimental_2010} and super-resolution schemes~\cite{ono_entanglement-enhanced_2013,tenne_super-resolution_2018}. Non-classical sources of light can also produce holograms~\cite{abouraddy_quantum_2001,asban_quantum_2019}, that were observed with single-photons~\cite{chrapkiewicz_hologram_2016-1} and photon pairs~\cite{devaux_quantum_2019}. \\Here, we introduce and experimentally demonstrate a holographic imaging concept that relies on quantum entanglement to carry the image information.
Phase images are encoded in the polarisation-entanglement of hyper-entangled photons and retrieved through spatial intensity correlation measurements (i.e. photon coincidence counting).%
This quantum holographic scheme has several distinguishing features: (i)  it is based on remote interferences between two distant photons, which removes the need for path overlap between the reference and illumination beams; (ii) it uses a subspace for encoding/decoding phase information that is robust against dephasing decoherence such as the presence of dynamic random phase disorder on the imaging paths; (iii) the reliance on a quantum illumination approach provides immunity to classical noise e.g. stray light falling on the sensor during measurement; (iv) entanglement enhances the spatial resolution by a factor $1.84$ compared to classical holography. \\
Finally, we demonstrate the potential of quantum holography beyond imaging, in particular for quantum state characterisation, by performing a spatially-resolved measurement of the Clauser-Horne-Shimony-Holt (CHSH) inequality to quantify hyper-entanglement in the generated quantum states.

\begin{figure*}
\includegraphics[width=0.9\textwidth]{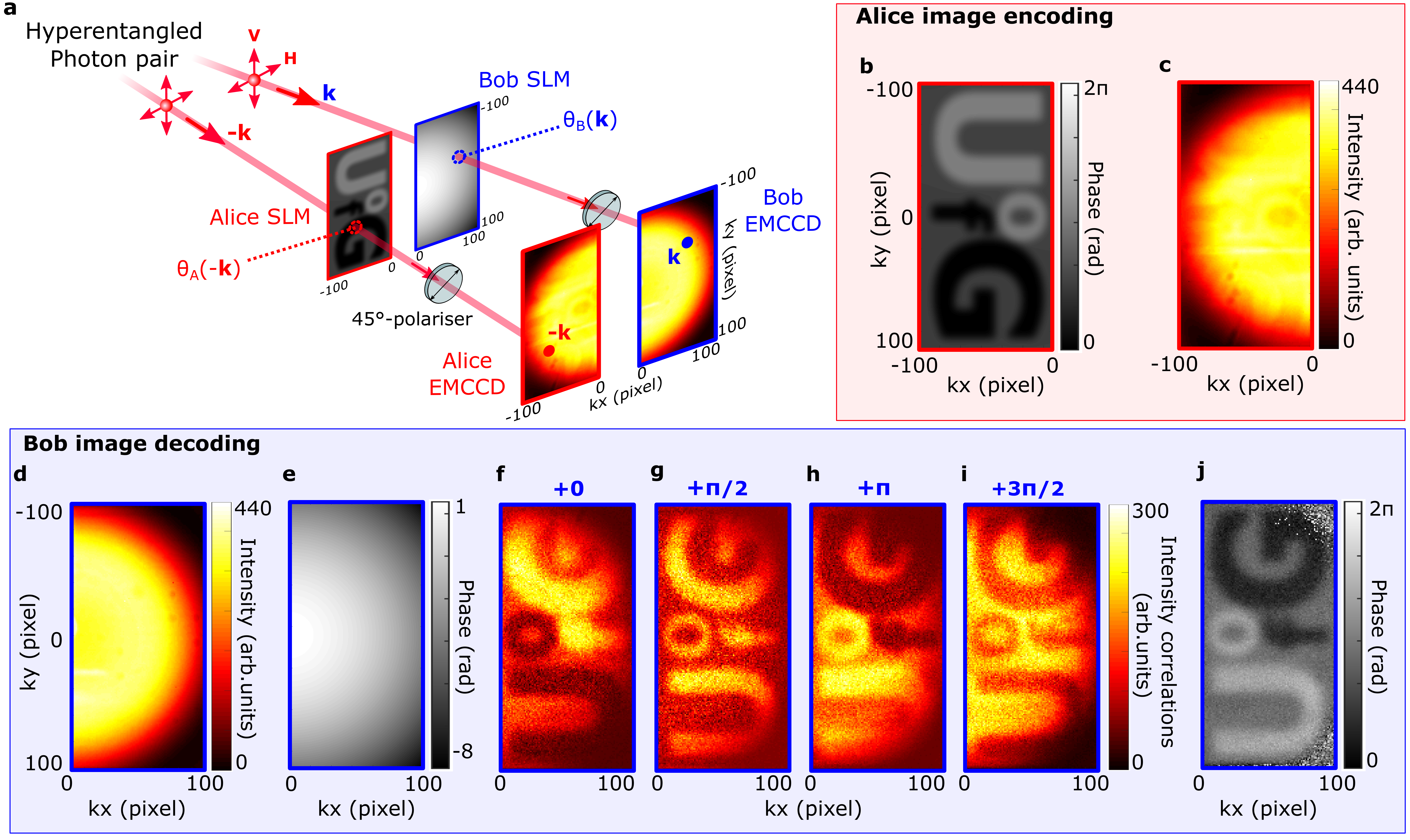} 
\caption{\label{Figure1} \textbf{Schematic of the quantum holographic reconstruction.} \textbf{a}, Space-polarisation hyperentangled photon pairs propagate through two spatial light modulators (Alice SLM and Bob SLM) and are detected by two electron multiplied charge couple device cameras (Alice EMCCD and Bob EMCCD). Transverse momentums $\vec{k}$ of photons with negative $x$-component $(k_x<0)$ are mapped to pixels on Alice's SLM and camera, while those with positive $x$-component $(k_x>0)$ are mapped to pixels on Bob's SLM and camera. Parallel aligned nematic liquid-crystal SLMs allow Alice and Bob to modulate at any pixel the horizontal polarisation of incoming photons with spatial phases $\theta_A$ and $\theta_B$. Two polarisers oriented at $45$ degrees are inserted between SLMs and cameras. \textbf{b}, Phase image $\theta_A(-\vec{k})$ displayed on Alice SLM. \textbf{c} and \textbf{d}, Intensity images measured by Alice and Bob on their cameras, respectively. \textbf{e}, SLM pattern displayed on Bob SLM to compensate for the static phase distortion $\Psi_0$. \textbf{f-i}, Intensity correlation images measured by Bob for different constant phase shift programmed on his SLM: $+0$ (\textbf{f}), $+ \pi/2$ (\textbf{g}), $+  \pi$ (\textbf{h}) and $+3 \pi/2$ (\textbf{i}). Each image is obtained by measuring intensity correlations between Bob camera pixels $\vec{k}$ and their symmetric on Alice camera $\vec{-k}$. \textbf{j}, Phase image reconstructed by Bob, with a signal-to-noise ratio (SNR) over $19$ and a normalised mean square error (NMSE) of $5 \%$. A total of $2.5 \times 10^6$ frames was acquired to retrieve the phase at a frame rate of $40$fps, which corresponds to $17$ hours of acquisition. Intensity and intensity correlation values are in arbitrary units and the same scales are used in all the figures of the manuscript. }
\end{figure*}

The conceptual arrangement of our quantum holographic scheme is illustrated in Fig.~\ref{Figure1}a. Photon pairs entangled in space and polarisation~\cite{barreiro_generation_2005} interact with two spatial light modulators (`Alice SLM' and `Bob SLM') and are then detected by two single-photon imaging devices, for example two distinct areas of an electron multiplied charge coupled device camera (`Alice EMCCD'  and `Bob EMCCD'). The transverse momentum $\vec{k}$ of the photons is mapped onto separated pixels of the SLMs and re-imaged onto the cameras. Alice and Bob shape and detect photons with momentum of negative $x$-component $(k_x<0)$ and positive $x$-component $(k_x>0)$, respectively. The quantum state of the photon pair after the SLMs is thus
\begin{equation}
\label{equ2}
\sum_{\vec{k}} \left[  \ket{V}_\vec{k}  \ket{V}_{-\vec{k}} + e^{i \Psi(\vec{k})}  \ket{H}_\vec{\vec{k}}  \ket{H}_{-\vec{k}}  \right]
\end{equation} 
in which $\Psi$ is a relative phase, $\ket{H}$ and $\ket{V}$ represent horizontal and vertical polarisation states of the photons. For a given momentum $\vec{k}$ $(k_x>0)$, $\Psi(\vec{k})$ is the sum of three phase terms $\Psi_0(\vec{k})$, $\theta_A(-\vec{k})$ and $\theta_B(\vec{k})$. $\Psi_0(\vec{k})$ is a static phase distortion produced during the photon generation process~\cite{hegazy_tunable_2015} that is characterised beforehand (see Methods). The phases $\theta_A(-\vec{k})$ and $\theta_B(\vec{k})$ are actively controlled by Alice and Bob by programming pixels at coordinates $-\vec{k}$ and $\vec{k}$ of their SLMs. This is made possible by the use of parallel aligned nematic liquid-crystal SLMs, that enable the manipulation of the horizontal polarisation of incoming photons but leave the vertical component unchanged.

\textbf{Quantum holography:}  First, Alice encodes an image $\theta_A(-\vec{k})$ in the phase component of entangled photons by programming her SLM with the corresponding phase pattern. Figure~\ref{Figure1}b shows the pattern used in our experiments, corresponding to the letters \textit{UofG}. On the other hand, Bob displays on his SLM a phase mask $\theta_B(\vec{k})=-\Psi_0(\vec{k})$ to compensate for the phase distortion $\Psi_0$ (Fig.~\ref{Figure1}e). This correcting phase remains superimposed to all phase masks that Bob programs throughout the experiment. As a result, the phase of the quantum state after the SLMs equals exactly the encoded image $\Psi(\vec{k})=\theta_A(-\vec{k})$. In the example shown in Fig.~\ref{Figure1}b, pixels associated with the letters \textit{U} and \textit{o} are encoded as the states $ \ket{V V}+\ket{H H}$ ($\Psi=0$), while \textit{f} and \textit{G} are encoded as $\ket{V V}-\ket{H H}$ ($\Psi=\pi$). After programming Alice's phase, we observe that the intensity images measured by both Alice and Bob, shown in Figs.~\ref{Figure1}c and e, are homogeneous and do not reveal the phase-encoded image. This observation remains valid when including polarisers in front of the cameras, at any orientation. 

In the holographic reconstruction step of the process, Bob decodes the image by performing intensity correlation measurements between pixels at $\vec{k}$ of his camera and symmetric pixels at $-\vec{k}$ on Alice's camera~\cite{defienne_general_2018-2}, with the two polarisers oriented at $45$ degrees (see Methods). This measurement is repeated four times for four different constant phase shifts $\theta$ applied on Bob's SLM, resulting in intensity correlation image $R_\theta(\vec{k}) \propto \, 1+\cos(\theta_A(\vec{k})+\theta)$ (see Methods). The phase image programmed on Alice's SLM must remain stationary during the full process, which takes up to several hours. Figures~\ref{Figure1}f-i show four intensity correlation images measured for $\theta \in \{0,\pi/2,\pi,3\pi/2 \}$ that partially reveal the hidden phase. Following a similar approach to classical holography, Bob then reconstructs the encoded image by using equation~\ref{equ1} after replacing $I_\theta$ by $R_\theta$. As shown in Fig.~\ref{Figure1}j, the retrieved image is $180$ degrees rotated and is of high quality, with a signal-to-noise ratio (SNR) over $19$ and a normalised mean square error (NMSE) of $5 \%$. While the SNR measures the intrinsic quality of the image retrieved by Bob in term of noise level, the NMSE quantifies its resemblance to the original image encoded by Alice (see Methods). Note that the combination of a highly multi-mode quantum state with a camera-based multi-pixel coincidence counting approach removes the need for operating raster-scanning approaches, as used for example in NOON state microscopy~\cite{ono_entanglement-enhanced_2013}.

\begin{figure*}
\includegraphics[width=0.6 \textwidth]{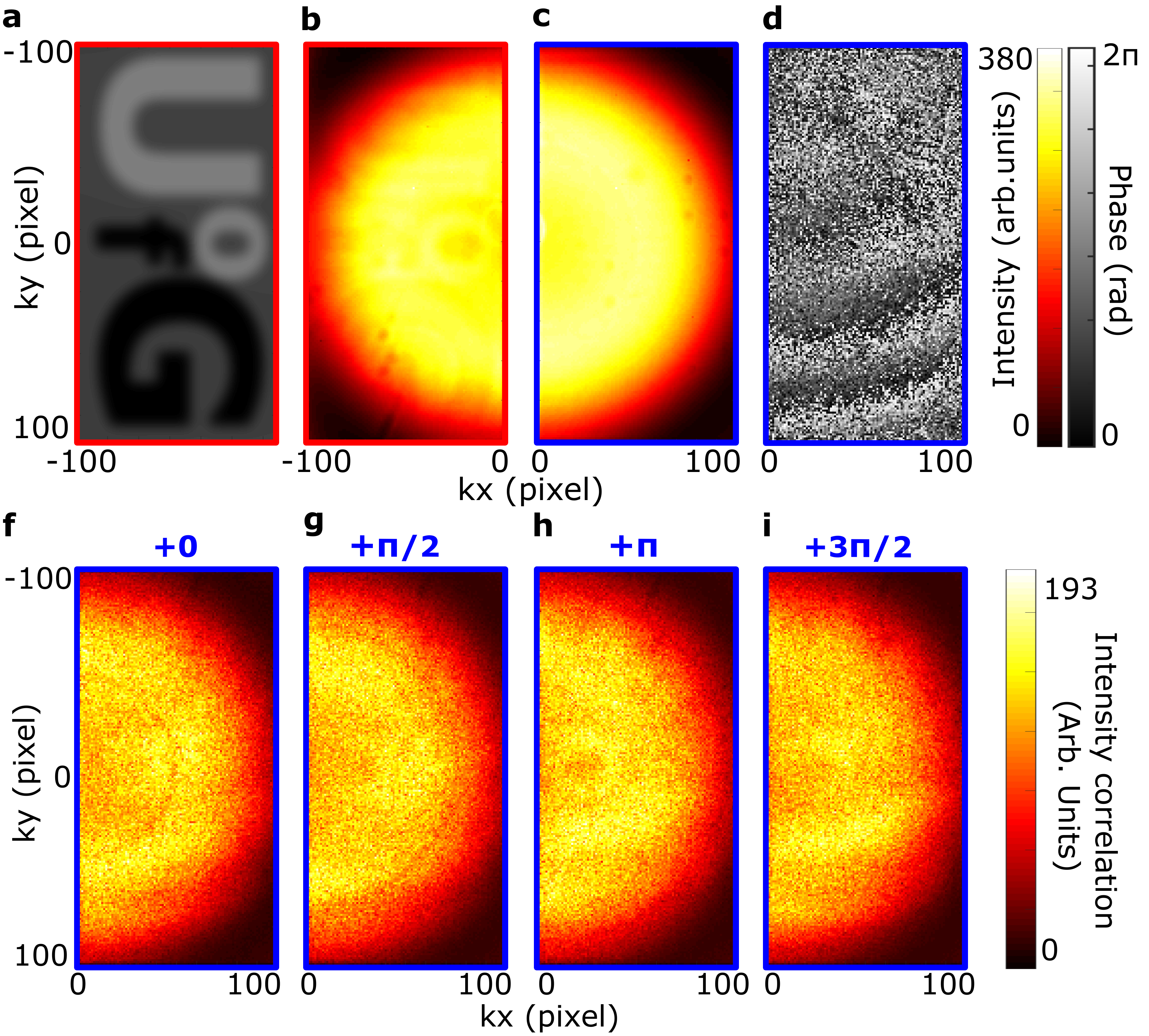} 
\caption{\label{Figure2}  \textbf{Quantum holography without polarisation entanglement}. \textbf{a}, Phase image encoded by Alice. \textbf{b} and \textbf{c}, Intensity images measured by Alice and Bob. \textbf{d}, Phase reconstructed by Bob that does not reveal the encoded image (NMSE=$95 \%$). \textbf{f-i}, Intensity correlation images used in the phase reconstruction process measured for different phase shifts: $+0$ (\textbf{f}), $+ \pi/2$ (\textbf{g}), $+ \pi$ (\textbf{h}) and $+3\pi/2$ (\textbf{i}). $2.5\times10^6$ frames were acquired in total. }
\end{figure*}

The photon pair spatial correlations provide the high-dimensional image space~\cite{howell_realization_2004} while polarisation entanglement carries the grey-scale information at each pixel. The presence of polarisation entanglement is therefore essential to this scheme. For example, Fig.~\ref{Figure2} shows results of quantum holography performed with the same encoded image as in Fig.~\ref{Figure1}a but using a source of photon pairs that are entangled in space but not in polarisation (see Methods). As in the previous case, intensity images measured by Alice and Bob in Fig.~\ref{Figure2}b and c do not reveal information about the encoded phase. However, Figs.~\ref{Figure2}f-i show that the intensity correlation images acquired during the phase-shifting process do not reveal any image information either, and the phase image cannot be retrieved (NMSE=$95 \%$), as shown in Fig.~\ref{Figure2}d. Non-zero values in intensity correlation images also confirm that (classical) correlations between photon polarisations are present without entanglement; the existence of a phase $\Psi$ is conditioned on the coherence between the two-qubit terms $\ket{VV}$ and $\ket{HH}$~\cite{kwiat_experimental_2001}, and thus on the entanglement in the state. However, this conclusion only concerns polarisation entanglement, because the presence of spatial entanglement is itself not strictly required in our holographic protocol. Therefore, one may design an experimental scheme using photons entangled in polarisation but only classically correlated in space to achieve similar results, even if this would be technically much more difficult than using hyper-entangled photons, with no real additional benefits.

\textbf{Robust subspaces and phase disorder:} Here the phase information is encoded and decoded from a subspace spanned by the two basis states $\ket{H}_{\vec{k}} \ket{H}_{\vec{-k}}$ and $\ket{V}_{\vec{k}} \ket{V}_{\vec{-k}}$. We verify that in our quantum holography concept, the use of this subspace protects the encoded phase information against dephasing decoherence generated by dynamic random phase disorders. Such robustness is linked to the notion of \textit{decoherence-free subspaces} (DFSs) that have been shown to protect quantum states against decoherence by exploiting symmetries in system-environment interactions~\cite{lidar_decoherence-free_2003}. Figure 3 describes an experimental apparatus in which space-polarisation entangled photons propagate through two thin diffusers (figure inset) positioned on a motorised translation stage in the image plane of both SLMs and cameras. In this configuration, the polarisation qubits at spatial mode $\vec{k}$ undergo the transformations $\ket{V}_{\vec{k}} \rightarrow e^{i \Phi(\vec{k})} \ket{V}_{\vec{k}}$ and $\ket{H}_{\vec{k}} \rightarrow e^{i \Phi(\vec{k})} \ket{H}_{\vec{k}}$, where $\Phi(\vec{k})=\Phi_H(\vec{k}) =\Phi_V(\vec{k})$ are the identical (i.e. phase disorder is non polarisation-sensitive) time varying random phase shifts in spatial mode $\vec{k}$ added on horizontal and vertical polarisations. The dynamic phase term $\Phi$ therefore factorises out and leaves the encoded phase $\Psi$ intact
\begin{equation}
\label{equ3}
\sum_{\vec{k}} e^{i \left[ \Phi(\vec{k})+ \Phi(-\vec{k}) \right]} \left[   \ket{V}_\vec{\vec{k}}  \ket{V}_{-\vec{k}}  + e^{i \Psi(\vec{k})} \ket{H}_\vec{k}  \ket{H}_{-\vec{k}} \right].
\end{equation} 
Figure~\ref{Figure4} shows the experimental reconstruction of a phase image through the dynamic phase disorders. Intensity images measured by Alice and Bob are shown in Figs.~\ref{Figure4}.b and c, respectively, and the image encoded by Alice (Fig.~\ref{Figure4}a) is very accurately reconstructed by Bob in Fig.~\ref{Figure4}d (SNR$=21$ and NMSE$=2\%$). Phase reconstruction is also achieved through a static phase disorder (see SI), confirming that such robustness does not originate from an averaging effect. Note that all entangled polarisation basis sets are robust in the presence of the generic random phase disorder considered here. The equivalent classical states encoded in the basis set $\{ \ket{H}_{\vec{k}}, \ket{V}_{\vec{-k}} \}$ would totally decohere under the same conditions. One may also show that our state is robust against other forms of decoherence with different symmetries, such as collective dephasing decoherence~\cite{kwiat_experimental_2000}, but that are less realistic in our experimental arrangement. These results show that the use of specific subspaces to encode information, as previously shown with DFSs for implementing robust quantum information processing protocols~\cite{kielpinski_decoherence-free_2001,viola_experimental_2001,yamamoto_robust_2008,banaszek_experimental_2004}, can also be useful in the context of imaging for reconstructing polarisation-sensitive phase objects through optical disorder. While certain classical common-path interferometers may achieve similar robustness, we underline that this is impossible in any classical non common-path interferometers because the presence of uncorrelated phase disorders in different arms would completely erase the classical phase information. Instead here, the coherence between the two interferometer arms, and thus the encoded phase, is preserved thanks to the presence of polarisation entanglement.

\begin{figure*}
\includegraphics[width=0.9 \textwidth]{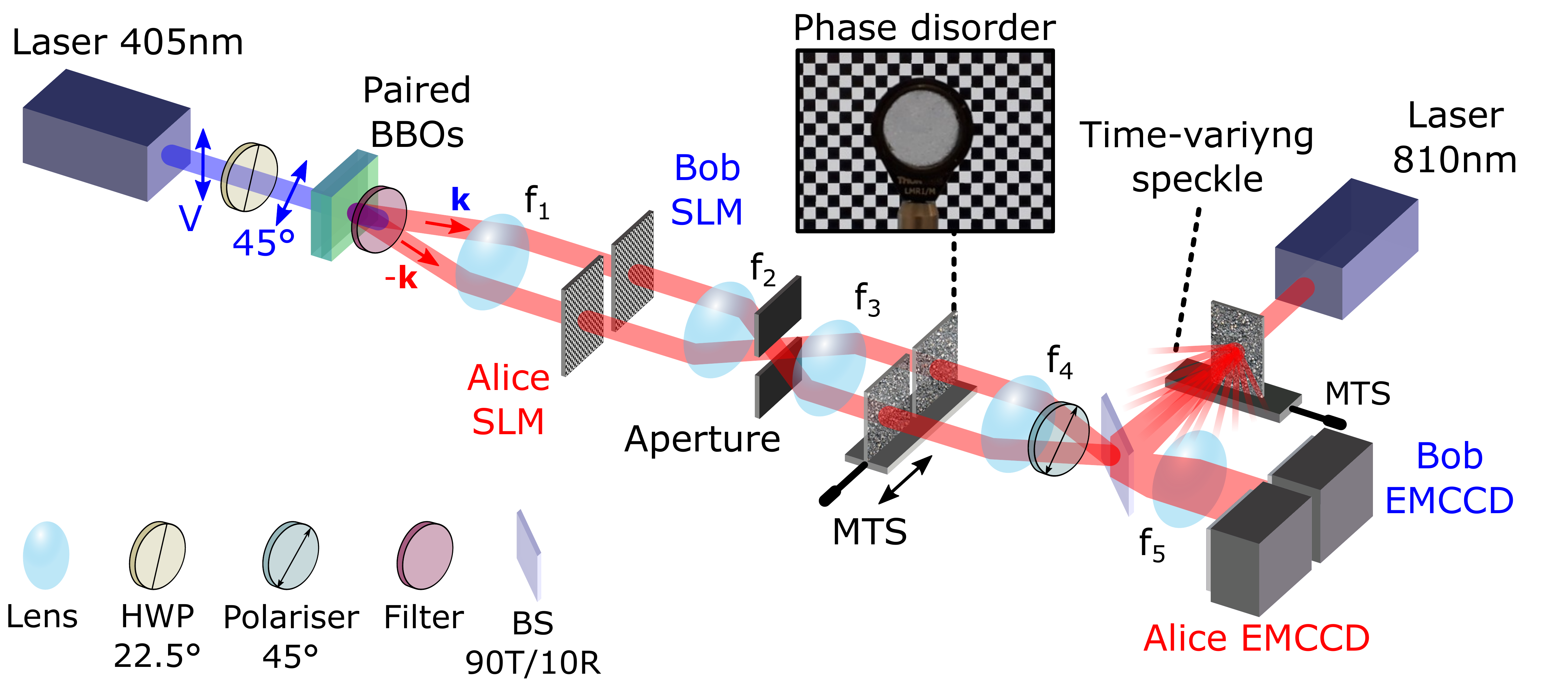} 
\caption{\label{Figure3}  \textbf{Experimental setup}. Light emitted by a laser diode at $405$nm and polarised at $45$ degrees illuminates a pair of $\beta$-Barium Borate (BBO) crystals (0.5mm thickness each) whose optical axes are perpendicular to each other to produce pairs of photons entangled in space and polarisation by type-I spontaneous parametric down-conversion (SPDC). After the crystals, pump photons are filtered out by a combination of long-pass and band-pass filters. The momentum of red photons is mapped onto an SLM divided in two parts (Alice SLM and Bob SLM) by Fourier imaging with lens $f_1$. Lenses $f_2-f_3$ image the SLM plane onto two thin diffusers (inset) positioned on a motorised translation stage (MTS) and lenses $f_4-f_5$ image it on an EMCCD camera split in two parts (Alice EMCCD and Bob EMCCD) i.e. each photon of a pair experiences a phase disorder independent of that experienced by its twin. A polariser at $45$ degrees is positioned between lenses $f_4$ and $f_5$. Stray light is inserted by illuminating another dynamic diffuser with a laser ($810$nm) to produce a time-varying speckle pattern that is superimposed on top of quantum light using an unbalanced beam splitter (BS 90T/10R). An adjustable aperture is positioned in the Fourier-plane of the SLMs. For clarity, only two propagation paths of entangled photons at $\vec{k}$ and $\vec{-k}$ are represented, while they have a higher dimensional spatial structure; SLMs, EMCCD cameras and diffusers are represented by pairs, while they are single devices spatially divided into two independent parts; the SLM is represented in transmission, while it operates in reflection. See Methods for further information.}
\end{figure*}

\begin{figure*}
\includegraphics[width=0.7 \textwidth]{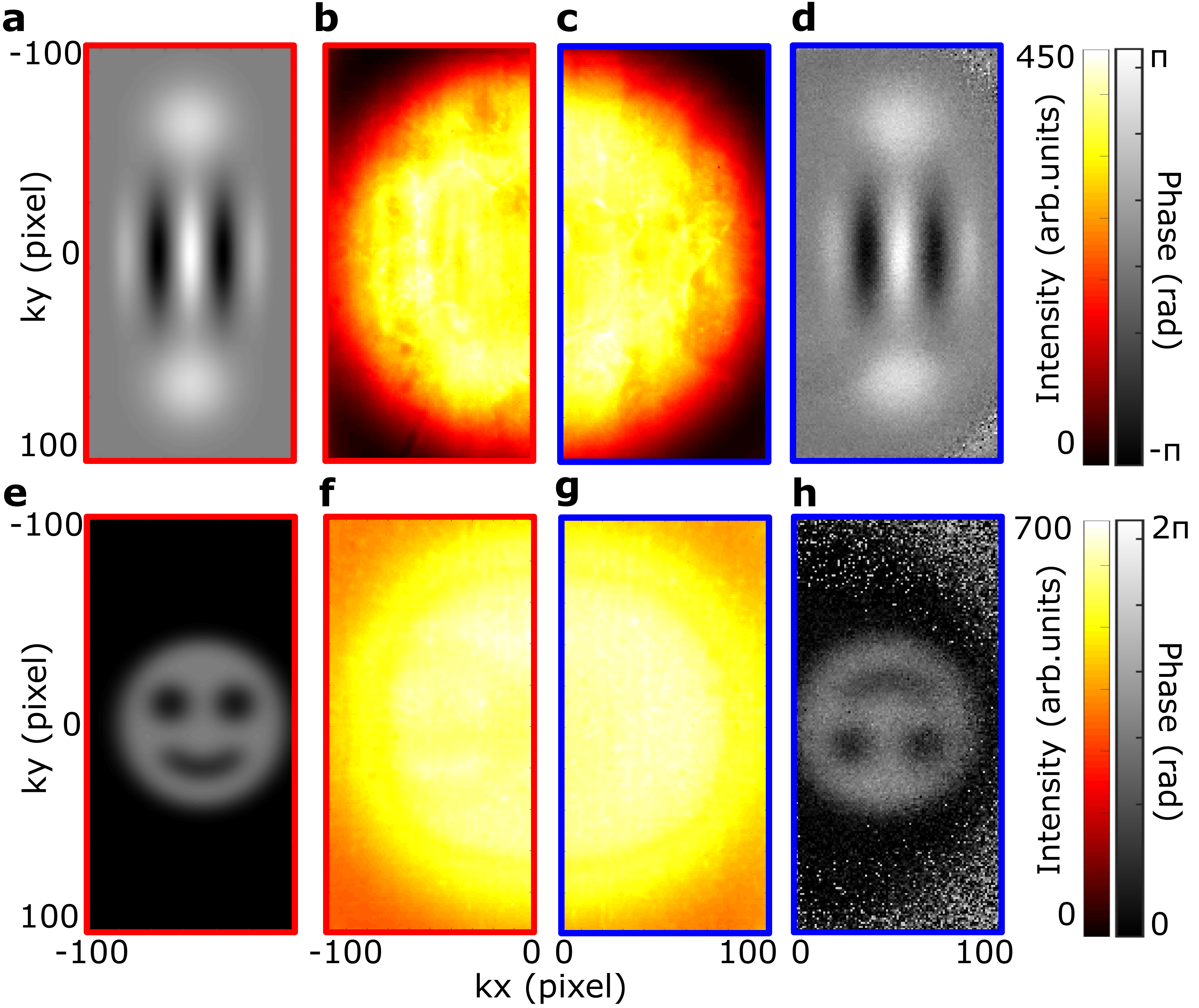} 
\caption{\label{Figure4} \textbf{Quantum holography through dynamic phase disorder and in the presence of stray light.} \textbf{a}, Phase image encoded by Alice. \textbf{b} and \textbf{c}, Intensity images measured by Alice and Bob through the dynamic phase disorder, respectively. \textbf{d}, Phase image reconstructed by Bob through the dynamic phase disorder with SNR=$21$ and NMSE=$2 \%$. \textbf{e}, Phase image programmed by Alice. \textbf{f} and \textbf{g}, Intensity images measured by Alice and Bob in the presence of dynamic stray light with an average intensity ratio classical/quantum of $0.5$, respectively. \textbf{h}, Phase image reconstructed by Bob in the presence of dynamic stray light with SNR=$9$ and NMSE=$17 \%$. All images were reconstructed from $5\times10^6$ frames.}
\end{figure*}

\textbf{Quantum illumination and dynamic stray light:} We have shown (e.g. Fig.~\ref{Figure1}) that phase information is reconstructed from a quantum illumination (QI) approach that relies on $4$ intensity correlation images obtained by coincidence counting~\cite{lloyd_enhanced_2008,tan_quantum_2008}. QI protocols use spatial correlation between photons to achieve enhanced imaging in the presence of noise, as recently demonstrated for amplitude objects illuminated by entangled pairs corrupted by \textit{static} stray light~\cite{defienne_quantum_2019,gregory_imaging_2020}. Here we exploit this robustness to image polarisation-sensitive phase objects in the presence of \textit{dynamic} stray light falling on both Alice and Bob sensors. Seen from the context of the QI proposal by S.Lloyd~\cite{lloyd_enhanced_2008}, the photon from the entangled pair detected by Bob plays the role of the 'ancilla' while its twin detected by Alice probes the object. As illustrated in Fig.~\ref{Figure3}, a time-varying speckle pattern is superimposed onto Alice and Bob sensors. This addition of classical light is clearly visible in the intensity images Figs.~\ref{Figure4}f and g. Because photons emitted by the classical source are not spatially correlated, they are not detected by intensity correlation measurements used for quantum phase reconstruction. Therefore, a phase image encoded by Alice (Fig.~\ref{Figure4}e) is accurately retrieved by Bob in the presence of dynamic classical stray light (Fig.~\ref{Figure4}h), with only a lower SNR compared to the case without stray light. Importantly, the SNR reduction does not indicate a permanent loss of image information content and can always be compensated by acquiring more frames (see SI for a quantitative analysis of SNR variation with quantum-classical intensity ratio and number of frames).

\begin{figure*}
\includegraphics[width=0.8 \textwidth]{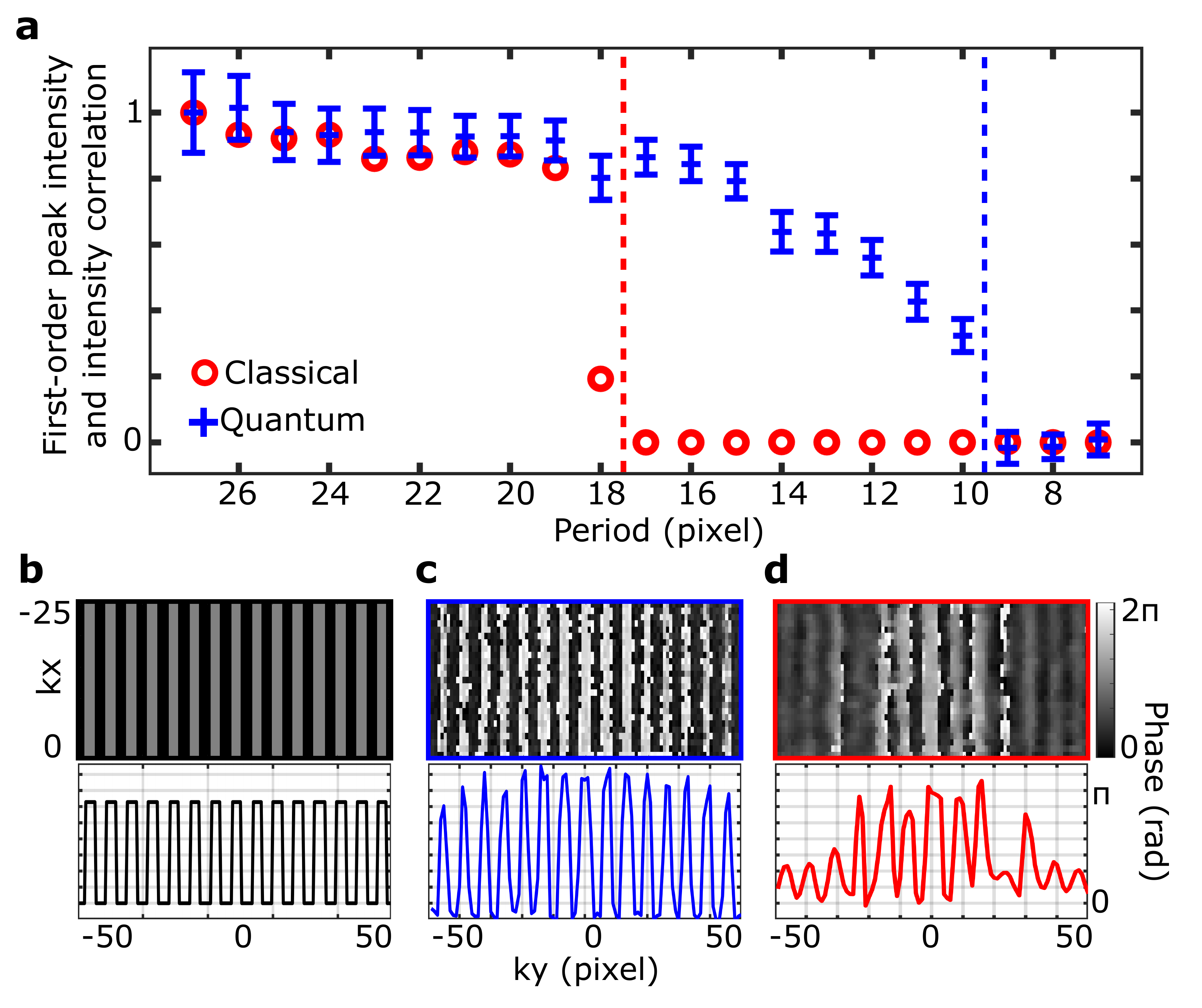} 
\caption{\label{Figure5} \textbf{Resolution enhancement.} \textbf{a}, Intensity (red circles, classical light) and intensity correlation values (blue crosses, quantum light) of first-order diffraction peaks measured for different grating periods in the presence of the aperture using the classical and quantum holographic systems, respectively. Cut-off periods are $17.5 \pm 0.5$ pixels and $9.5 \pm 0.5$. \textbf{b}, $0-\pi$ phase grating with $16$-pixels-period encoded by Alice (zoom $25 \times 100$ pixels).  \textbf{c} and \textbf{d}, Phase images of a $16$-pixels-period phase grating reconstructed using classical and quantum holographic systems (zoom $25 \times 100$ pixels), respectively. Corresponding projections along the $k_x$ axis are shown bellow (black, red and blue curves). $5 \times 10^6$ frames were acquired in total for each case under quantum illumination. Error bars of red points are too small to be shown on the graph.}
\end{figure*}

\textbf{Entanglement and spatial resolution:} Resolution enhancement using entangled photon pairs has been theorised~\cite{giovannetti_sub-rayleigh-diffraction-bound_2009} and exploited in \textit{scanning-based} imaging approaches~\cite{boto_quantum_2000,mitchell_super-resolving_2004}. This effect lies in the foundations of optical coherence, precisely reflecting the difference between its first and second order degrees~\cite{glauber_quantum_1963}. We demonstrate it here in the context of \textit{full-field} quantum holographic imaging, using a classical coherent holographic imaging system for comparison (details in Methods). We insert an aperture in the Fourier plane of the SLMs (Fig.~\ref{Figure3}) to control the transmitted spatial frequencies and we then image this plane onto the camera by replacing the lens $f_5$ by a lens of half-focal length (i.e. single-lens imaging). Phase grating objects with different periods are then programmed onto Alice SLM. When using classical holography, we observe in Fig.~\ref{Figure5}a that the intensity of the first-order diffraction peak vanishes for grating periods below $17.5 \pm 0.5$ pixels. Conversely, using our quantum holographic system, the first-order diffraction peak of intensity correlation diffraction pattern only disappears for a shorter period of $9.5 \pm 0.5$ pixels (see Methods and SI). The difference in frequency cut-off between the two systems corresponds to an enhancement of the spatial resolution by a factor $1.84 \pm 0.05$~\cite{goodman_introduction_2005}, close to the maximum theoretical value of $2$~\cite{giovannetti_sub-rayleigh-diffraction-bound_2009}. The resolution enhancement effect is also observed using the imaging configuration (i.e. with lens $f_5$): a $16$-pixels-period phase grating can be near-perfectly resolved using our quantum holographic approach (Fig.~\ref{Figure5}c) while a significant degradation is observed when using classical holography (Fig.~\ref{Figure5}d).

\begin{figure*}
\includegraphics[width=0.5 \textwidth]{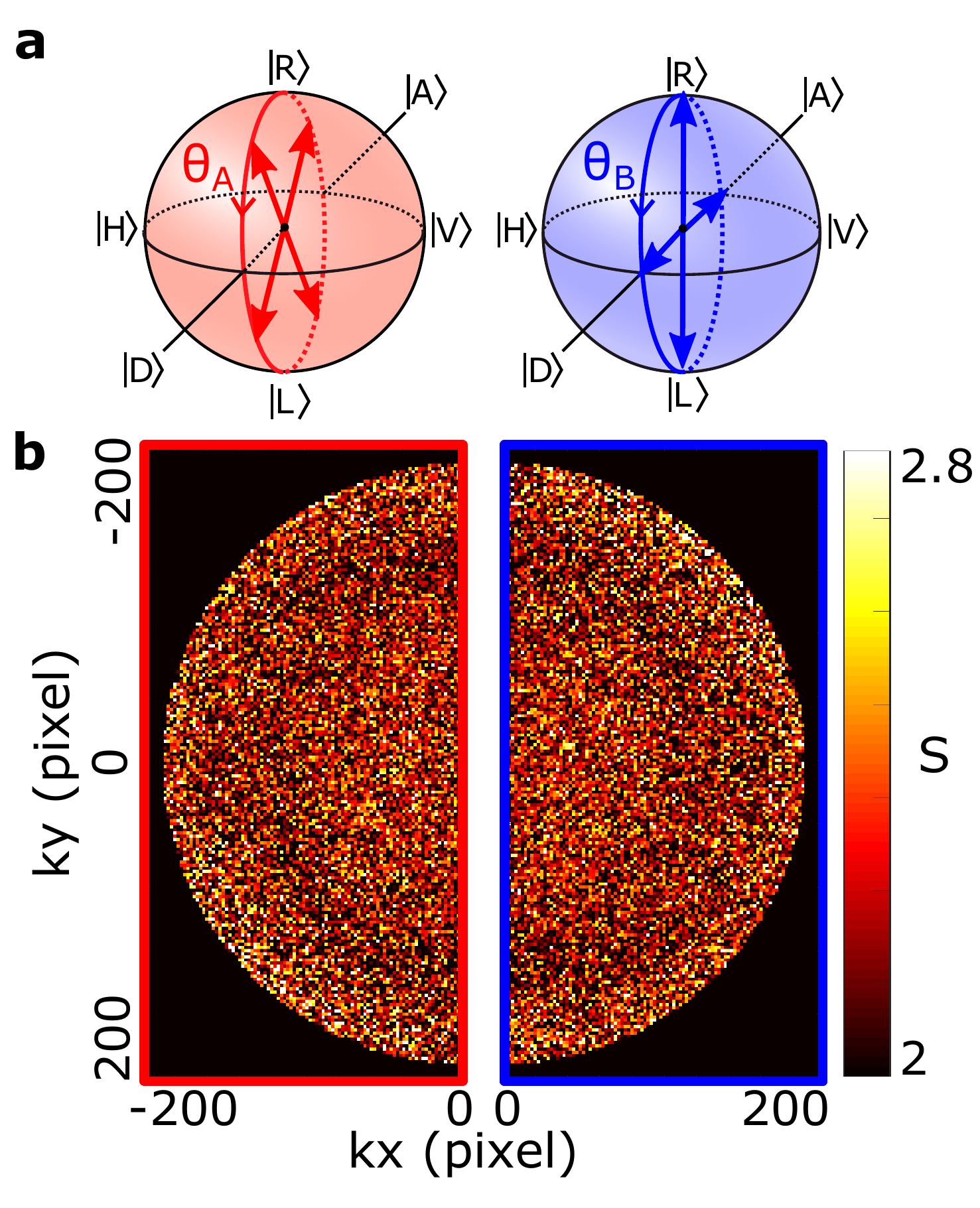} 
\caption{\label{Figure6} \textbf{Spatially-resolved Clauser-Horne-Shimony-Holt (CHSH) inequality violation.} \textbf{a}, Poincar\'{e} spheres representing the projections performed by Alice and Bob with their SLMs. $\ket{D}$ and $\ket{A}$ are two eigenstates of the diagonal polarisation basis and $\ket{R}$ and $\ket{L}$ are the two eigenstates of the circular polarisation basis. Bob performs measurements directly in the diagonal and circular basis, while Alice operates in a similar basis rotated by $\pi/4$. \textbf{b,} Values of $S$ measured at different pairs of pixels of Alice's and Bob's sensors within two half-disks containing $14129$ pairs of pixel. The two half-disks correspond to areas where the direct intensity is non-null and homogeneous. Images are calculated from intensity correlation measured for $16$ combinations of angles $\theta_A$ and $\theta_B$ (shown in Figure B of the SI). CHSH inequality is violated ($S>2$) in $10789$ pairs of pixels over the total of $14129$. An average value $\langle S\rangle=2.20 \pm 0.003 > 2$ is obtained by averaging $S$ values over the full two half-disks areas. See also Methods for further details. }
\end{figure*}

\textbf{Hyper-entanglement characterisation in high dimensions:} By harnessing the fundamental link between quantum state tomography and second-order optical coherence holography, the quantum holography scheme enables characterisation of hyper-entanglement in high dimensions. Indeed, the measurements performed by Bob in the phase-stepping holographic process correspond to projections in the diagonal ($\theta_B \in \{ 0,\pi \}$) and circular ($\theta_B \in \{ \pi/2,3\pi/2 \}$) polarisation basis (Figure~\ref{Figure6}a). Similarly, Alice can use her SLM to perform measurements in the corresponding rotated basis $\theta_A \in \{\pi/4,3\pi/4,5\pi/4,7\pi/4 \}$. During this process, the use of a compensation phase mask programmed on Bob SLM (Figure~\ref{Figure1}.e) is important to ensure an optimal orientation of the measurement bases. As shown in Fig.~\ref{Figure5}b, these measurement settings provide spatially-resolved measurements CHSH inequality across $10789$ pairs of pixels. Taking into account the finite momentum correlation width of the photons $\sigma_k =  [1.326 \pm 0.001]\times10^3$ rad.m$^{-1}$ ($\sim 1.1$ pixel), one may conclude that Alice and Bob share up to $8900$ polarisation-entangled states in parallel.Furthermore, an additional measurement in the position-space of the photons enables to measure their position correlation width $\sigma_r = 10.85 \pm 0.06$ $\upmu$m~\cite{moreau_realization_2012,edgar_imaging_2012}. One can then also verify the presence of spatial entanglement in the quantum state through an Einstein-Podolsky-Rosen (EPR) type inequality~\cite{howell_realization_2004}, which in our case results in $ \sigma_r \sigma_k = [1.44 \pm 0.01]\times10^{-2} < \frac{1}{2}$  (see Methods). 

\textbf{Application potential:} Our quantum holography concept can be applied also to the imaging of real-world objects, aside from the phase patterns imprinted on the SLM shown so far. As an example, in the SI we show phase images obtained by removing the SLM and therefore passing the entangled photons through bird feathers and adhesive tape. In biology, our scheme can be useful for measuring small variations of birefringence in biological structures for investigating cell pathophysiology~\cite{lee_quantitative_2013}, tissue damage~\cite{wang_birefringence_2018} and for ophthalmologic preclinical diagnosis~\cite{pircher_imaging_2004,gotzinger_measurement_2004}, typical situations in which classical holographic techniques can be limited by the presence of specimen-induced phase distortions (i.e. phase disorder) and stray light that cannot be blocked. Furthermore, combining our approach with concepts from Differential Interference Contrast imaging~\cite{allen_zeiss-nomarski_1969} enables to extend its range of applications to non polarisation-sensitive phase objects, with the potential for large field-of-view imaging microscopy~\cite{terborg_ultrasensitive_2016}. An illustration of such an extended setup is shown in Figure O of the SI, together with experimental results of quantum phase imaging of non-birefringent silicone oil droplets. Beyond the optical domain, our approach could also be extended to other imaging methods such as electron-based techniques~\cite{madan_holographic_2019} with potential for investigating complex biological systems at low radiation and further enhanced resolution. \\
Quantum states entangled in high dimensions and multiple degrees of freedom are also promising for moving beyond the limitations of current quantum communication and information processing technologies~\cite{erhard_advances_2020,barreiro_beating_2008,graham_superdense_2015,deng_quantum_2017-1}. One of the central challenges is ascertaining the presence of entanglement in a given quantum state, however complex and large it may be. In this respect, our quantum holography concept can be used for characterising entanglement in both space and polarisation distributed across up to $10^4$ modes, a task that would be prohibitively time consuming (if not impossible) using raster-scanning and single-outcome projective measurement techniques~\cite{barbieri_polarization-momentum_2005,krenn_generation_2014}. 

\textbf{Conclusions.} In summary, we have shown that it is possible to perform holography without (first order) coherence, a concept that is not possible in classical physics and broadens the remit of what is achievable with holography. Holographic imaging is enabled by quantum entanglement that indeed does not rely on classical optical coherence. Information about the image is encoded in the relative phases between the polarisation entangled two-photon qubits states (i.e. phase $\Psi$ in $\ket{VV}+e^{i \Psi} \ket{HH}$) and is distributed over the transverse spatial dimension through the high-dimensional structure of spatial entanglement. By harnessing the physical concepts linked to the notion of entanglement, including QI and DFS, it is possible to achieve resolution-enhanced measurement of polarisation-sensitive phase objects through random phase disorder and stray light, with practical advantages over classical holography. Furthermore, there is a fundamental correspondence between quantum holography and quantum tomography, that extends the concept to quantum state characterisation, including the analysis of hyper-entangled states in high-dimensions. One current practical limitation of our quantum holographic protocol is its long acquisition time (i.e. on the order of several hours) resulting from the low frame rate of EMCCD cameras. However, thanks to the rapid development of faster and cheaper sensors for imaging quantum correlations~\cite{lubin_quantum_2019,ndagano_imaging_2020}, we expect quantum holography to move towards practical applications for biological imaging and sensing, but also for characterising complex high-dimensional quantum states, that are likely to be at the heart of tomorrow's quantum optical communications and information processing technologies.


\newpage

\clearpage
\section*{Methods} 

\noindent \textbf{Experimental layout.} A paired set of BBO crystals have dimensions of $0.5 \times 5 \times 5$ mm each and are cut for type I SPDC at $405$ nm. They are optically contacted with one crystal rotated by $90$ degrees about the axis normal to the incidence face. Both crystals are slightly rotated around horizontal and vertical axis to ensure near-collinear phase matching of photons at the output (i.e. rings collapsed into disks). The pump is a continuous-wave laser at $405$ nm (Coherent OBIS-LX) with an output power of approximately $200$ mW and a beam diameter of $0.8\pm0.1$ mm. A $650$ nm-cut-off long-pass filter is used to block pump photons after the crystals, together with a band-pass filter centred at $810 \pm 5$ nm. The SLM is a phase only modulator (Holoeye Pluto-2-NIR-015) with $1920 \times 1080$ pixels and a $8$ $\upmu$m pixel pitch. The camera is an EMCCD (Andor Ixon Ultra 897) that operates at $-60^{\circ}$C, with a horizontal pixel shift readout rate of $17$ MHz, a vertical pixel shift every $0.3$ $\upmu$s, a vertical clock amplitude voltage of $+4$V above the factory setting and an amplification gain set to $1000$. It has a $16 \upmu$m pixel pitch. Exposure time is set to $3$ ms. The camera speed is about $40$ frame per second when considering a region of interest of $200 \times 200$ pixels, allowing to retrieved a phase image with SNR$\sim 20$ in about $17$ hours (i.e. $2.5.10^6$ frames in total). The characterisation of our system shows that the camera detects approximately $0.1$ pairs per spatial mode per second. The detection efficiency of the entire setup is approximately $0.48$. The classical source is a superluminescent diode laser (Qphotonics) with a spectrum of $810 \pm 15$ nm that is filtered using a band-pass filter at $810 \pm 5$ nm to match the photon pair's spectrum. The lens $f_1$ is composed by a series of three lenses of focal lengths $45$ mm - $125$mm - $150$ mm positioned into a Fourier imaging configuration i.e.the distance between each pair of lenses equals the sum of the focal lengths of each lens. The whole system can be seen as an lens $f_1$ of effective focal length $f_1 = 54$ mm. Focal lengths of the other lenses are $f_2 = 150$ mm, $f_3 = f_4 = 75$ mm, $f_5 = 100$ mm and $f_6 = 175$ mm. Distances: crystal plane - lens $f_1$ $= 54$mm ; lens $f_1$ - SLMs $= 54$mm ;  SLMs - lens $f_2$ $=150$mm ; lens $f_2$ - aperture $=150$mm ; aperture - lens $f_3$ $=75$mm ; lens $f_3$ - phase disorder $=75$mm ; phase disorder $=75$mm - lens $f_4$ $=75$mm ; lens $f_4$ - beam splitter $=75$mm ; beam splitter - lens $f5$ $=100$mm ; lens $f5$ - camera $=100$mm. The magnification factor from the SLM plane to the camera plane is $0.7$. The diffuser is a plastic sleeve layer of thickness $<100$ $\upmu$m, roughness of $46$ $\upmu$m and has a decorrelation time of $183$ ms. See the SI section 6 and 9 for further details on the diffuser properties and the quantum source.  \\
 
\noindent \textbf{Intensity correlation images.} The camera sensor is split in two identical regions of interest composed of $201 \times 101$ pixels associated with Alice and Bob. To measure intensity correlations, the camera first acquires a set of $N$ images. Then, values of intensity correlation $R(\vec{k})$ between pixel at $\vec{k}$ on Bob's side and the symmetric at $\vec{-k}$ on Alice's side are calculated by subtracting the product of intensity values measured in the same frame by the product of intensity values measured in successive frames, and averaging over all the frames: 
\begin{equation}
\label{equ4}
R(\vec{k}) = \frac{1}{N} \sum_{l=1}^N \left[ I_l(\vec{k})I_l(\vec{-k})-I_l(\vec{k})I_{l+1}(\vec{-k}) \right].
\end{equation}
in which $I_l$ denotes the $l^{th}$ frame~\cite{defienne_general_2018-2}. See the SI section 1 for further details on the intensity correlation measurement. 

\noindent \textbf{Quantum holography.} Intensity correlation measurement performed between pixels $\vec{k}$ and $\vec{-k}$ with two polarisers oriented at $45$ degrees positioned in front of the cameras can be associated with the following measurement operator:
\begin{eqnarray}
\label{equ5}
&& \frac{1}{2} \big[ \ket{H}  \bra{H}  + \ket{V}  \bra{V} +\ket{H}  \bra{V}  + \ket{V}  \bra{H} \big]_\vec{k} \nonumber \\
&& \otimes \, \big[ \ket{H}  \bra{H}  + \ket{V}  \bra{V} +\ket{H}  \bra{V}  + \ket{V}  \bra{H} \big]_\vec{-k} 
\end{eqnarray}
For a given pair of pixels ($\vec{-k},\vec{k}$), the expectation value of this operator in the state described by equation~\ref{equ2} is
\begin{equation}
R(\vec{k}) = \frac{1}{2} \left[ 1 + \cos(\Psi(\vec{k})) \right]
 \end{equation}
During the holographic process, Alice encodes a phase $\theta_A(-\vec{k})$ and Bob applies a phase shift $\theta$ superimposed over the phase compensation pattern $-\Psi_0(\vec{k})$. As a result, intensity correlation measurements performed by Bob for a given $\theta$ are given by $R_\theta(\vec{k}) = \frac{1}{2} [ 1+\cos(\theta_A(\vec{k})+\theta)]$. As in classical holography (equation~\ref{equ1}), Bob then reconstructs the phase image $\theta_A(\vec{k})$ image using four successive measurements: $\theta_A(\vec{k}) = \arg \left[ R_0(\vec{k})-R_\pi(\vec{k}) + i \left( R_{\pi/2}(\vec{k})-R_{3\pi/2}(\vec{k}) \right) \right]$. Note that, to take into account a more general case, the state in equation~\ref{equ2} can be re-written as:
\begin{equation}
\label{equ6}
\sum_{\vec{k}} \left[ e^{i \Psi(\vec{k})}  \ket{H}_\vec{\vec{k}}  \ket{H}_{-\vec{k}}  + \alpha \ket{V}_\vec{k}  \ket{V}_{-\vec{k}} \right]
\end{equation}   
with $\alpha \in ]0,1]$. In this case, the expectation value of the operator in equation~\ref{equ5} changes into $\frac{1}{2} \left[ 1 + \alpha^2 \cos(\Psi(\vec{k})) \right]$, but $\theta_A(\vec{k})$ is still retrieved using equation~\ref{equ1} (albeit with visibility equal to $\alpha^2$).\\

\noindent \textbf{Characterisation of spatial entanglement.}
Spatial entanglement in the photon source is characterised by performing intensity correlation measurements between positions and momentum of photons, using the method described in~\cite{moreau_realization_2012,edgar_imaging_2012}. Correlation width measurements return values of $\sigma_r = 10.85 \pm 0.06$ $\upmu$m for position and $\sigma_k = [1.326 \pm 0.001]\times10^3$rad.m$^{-1}$ for momentum. These values show violation of EPR criteria $ \sigma_r \sigma_k = [1.44 \pm 0.01]\times10^{-2} < \frac{1}{2}$~ \cite{howell_realization_2004}. See the SI section 2 for further details, including the correlations images in position and momentum spaces used to estimate the correlation widths. \\

\noindent \textbf{Phase distortion characterisation.}
The phase distortion $\Psi_0(\vec{k})$ originates from the birefringence of the paired BBO crystals used to generate photon pairs~\cite{hegazy_tunable_2015}. $\Psi_0(\vec{k})$ is measured beforehand by performing a holographic measurement between a flat phase pattern programmed on Alice SLM and successive phase shifts displayed on Bob SLM. This characterisation process results in a phase distortion of the form $\Psi_0(k_x,k_y) = 4.69 k_x^2+5.04 k_y^2+0.02$. In our experiment, a correcting phase mask is directly programmed on Bob SLM to compensate for the phase distortion (Fig.~\ref{Figure1}.e). For holographic imaging of phase objects, we note that in principle it would be possible to replace Bob SLM by a rotating polariser positioned in front of the camera and compensate for the phase distortion afterwards in a post-processing step on a computer. However, the use of a correcting pattern directly implemented on Bob's SLM is important for performing the spatially resolved CHSH measurement (Fig.~\ref{Figure6}) because it ensures an optimal orientation of the measurement bases.. See the SI section 4 for further details on the phase distortion characterisation.  \\

\noindent \textbf{Signal-to-noise, normalised mean square error and spatial resolution.}
Signal-to-noise ratio (SNR) is obtained by calculating an averaged value of the phase in a region of the retrieved image where it is constant, and then dividing it by the standard deviation of the noise in the same region. To have a common reference, SNR values are calculated using areas where the phase is constant and equals $\pi$. For a fixed exposure time and pump power, the SNR varies as $\sqrt{N}$, where $N$ is the number of images used to reconstruct the intensity correlation images~\cite{reichert_optimizing_2018}. In the presence of stray light, the SNR decreases as $1/\langle I_{cl} \rangle$, where $\langle I_{cl} \rangle$ is the average intensity of classical light falling on the sensor~\cite{defienne_quantum_2019,gregory_imaging_2020}. Quantitative analysis of the SNR variation with the number of frames and the intensity of stray light are provided in sections 3 and 7 of the SI.\\
The normalised mean square error (NMSE)~\cite{gonzalez_digital_1977} quantifies the resemblance between an image reconstructed by Bob and the ground truth image encoded by Alice. The NMSE is calculated using the formula:
\begin{equation}
NMSE= \frac{M_0}{M_\infty}
\end{equation}
where $M_0$ is the mean square error (MSE) measured between the ground truth and the retrieved image and $M_\infty$ is an average value of MSE measured between the ground truth and a set of images composed of phase values randomly distributed between $0$ and $2 \pi$. The MSE between two images composed of $P$ pixels with values denoted respectively $\{x_i\}_{i \in  [\![ 1, P ]\!]}$ and $\{y_j\}_{j \in  [\![ 1, P ]\!]}$ is defined as $M = 1/P \sum_{i=1}^P |x_i-y_i|^2$. Values of NMSE range between $1$ (retrieved image is a random phase image) and $0$ (retrieved image is exactly the ground truth). Spatial resolution in the retrieved image is determined by the spatial correlation width of entangled photons. In our experiment, its value is estimated to $d = 45 \pm 3$ $\upmu$m, which corresponds to approximately $3$ camera pixels. See the SI section 3 for further details on the spatial resolution characterisation.\\

\noindent \textbf{Photons without polarisation entanglement.}
Results shown in Fig.~\ref{Figure2} are obtained using a quantum state defined by the following density operator:
\begin{equation}
\frac{1}{2} \sum_{\vec{k}} \big[ \ket{H}_\vec{\vec{k}}  \ket{H}_{-\vec{k}} \bra{H}_\vec{\vec{k}}  \bra{H}_{-\vec{k}}  +  \ket{V}_\vec{k}  \ket{V}_{-\vec{k}} \bra{V}_\vec{k}  \bra{V}_{-\vec{k}} \big]
\end{equation}
Experimentally, it is produced by switching the polarisation of the pump laser between vertical and horizontal polarisations, which is equivalent of using a unpolarised pump. Because entanglement originates fundamentally from a transfer of coherence properties between the pump and the down-converted fields in SPDC~\cite{jha_spatial_2010,kulkarni_intrinsic_2016,kulkarni_transfer_2017}, the lack of coherence in the pump polarisation induces the absence of polarisation entanglement in the produced two-photon state, while spatial and temporal entanglement are maintained~\cite{kwiat_experimental_2001}. See the SI section 5 for further details on state entangled in space but not in polarisation.\\

\noindent \textbf{Reference classical holographic system.}
Experimental results shown in Figure~\ref{Figure5} were obtained using a holographic system that is a classical version of our quantum protocol, namely a polarisation phase-shifting common-path holographic interferometer~\cite{mukhopadhyay_polarization_2013}. In this classical system, a collimated laser beam ($810$nm) polarised at $45$ degrees illuminates Alice SLM on which a phase object is programmed (Bob SLM is not used). Alice SLM is imaged onto a single EMCCD camera using the same imaging system as the one described in Figure~\ref{Figure3}. Phase-shifting holography is then performed by superimposing four constant phase patterns ($0,\pi/2,\pi,3 \pi/2$) on top of the programmed phase object and measuring the four corresponding intensity images on the camera. Finally, the phase object is reconstructed using equation~\ref{equ1}. See the SI section 8 for more details.\\

\noindent \textbf{Resolution enhancement measurement.} A comparison of spatial resolution between quantum and classical holographic systems is performed by measuring their respective frequency cut-off~\cite{goodman_introduction_2005}. Results shown in Figs.~\ref{Figure5} are obtained by replacing the lens $f_5$ in Fig.~\ref{Figure3} by a lens with half-focal length ($f_5/2 = 50$mm) to directly image the Fourier plane of the SLM onto the EMCCD camera. The aperture is also placed in this Fourier plane. In the classical case, measurements are performed by illuminating Alice SLM with a $45$-degrees collimated laser beam (see previous Methods section). When programming a phase grating on the SLM, intensity images measured by the EMCCD shows a diffraction pattern with three main components: a central zero-order peak and two symmetrically positioned plus-or-minus first-order peaks. Red circles in Fig.~\ref{Figure5}.a correspond to the intensity of the (plus) first-order peak measured for different grating periods. Because of the aperture, a sharp cut-off is observed at period of $17.5 \pm 0.5$ pixels. In the quantum case, diffraction patterns are revealed by measuring intensity correlations with the EMCCD camera (i.e. second-order coherence)~\cite{devaux_quantum_2019,asban_quantum_2019}. More precisely, an complete intensity correlation matrix $R(\vec{r_1},\vec{r_2})$ is measured for each phase grating using a generalised version of equation~\ref{equ4}~\cite{defienne_general_2018-2}:
\begin{equation}
R(\vec{r_1},\vec{r_2}) = \frac{1}{N} \sum_{l=1}^N \left[ I_l(\vec{r_1})I_l(\vec{r_2})-I_l(\vec{r_1})I_{l+1}(\vec{r_2}) \right]
\end{equation}
where $N$ is the number of acquired frames, $\vec{r_1}$ and $\vec{r_2}$ are spatial positions in the Fourier plane (i.e. camera pixel positions). Then, the intensity correlation matrix is projected along the minus-coordinate axis $\vec{\delta r} = \vec{r_1}-\vec{r_2}$ using the formula:
\begin{equation}
P(\vec{\delta r}) = \sum_{\vec{r}} R(\vec{r},\vec{r}-\vec{\delta r}) .
\end{equation}
where the summation is performed over all illuminated pixels $\vec{r}$. The use of such projection to reveal diffraction patterns under quantum illumination was demonstrated in~\cite{defienne_adaptive_2018-3,devaux_quantum_2019}. Similarly to the classical case, three peaks of intensity correlations are observed when visualizing intensity correlation in the minus-coordinate basis (diffraction patterns are shown in Figure K of the SI). Blue crosses in Fig.~\ref{Figure5}.a correspond to the intensity of the (plus) first-order peak measured for different grating periods. In this case, a cut-off is observed at period of $9.5 \pm 0.5$ pixels, which corresponds to a resolution enhancement of $17.5/9.5 = 1.84 \pm 0.05$. See the SI section 8 for further details on the resolution characterisation, including detailed experimental schemes and images of classical and quantum diffraction patterns.\\

\noindent \textbf{Clauser-Horne-Shimony-Holt (CHSH) measurement.} A set of $16$ intensity correlations images $R_{\theta_A,\theta_B}$ is first measured using all combinations of uniform phases $\theta_A \in \{\pi/4,3\pi/4,5\pi/4,7\pi/4 \}$ and $\theta_B \in \{0,\pi/2,\pi,3\pi/2 \}$ programmed on Alice and Bob SLMs. Then, a correlation image $E_{\theta_A,\theta_B}$ is calculated using the following formula~\cite{clauser_proposed_1969}:
\begin{equation}
E_{\theta_A,\theta_B} = \frac{R_{\theta_A,\theta_B}-R_{\theta_A,\theta_B+\pi}-R_{\theta_A+\pi,\theta_B}+R_{\theta_A+\pi,\theta_B+\pi}}{R_{\theta_A,\theta_B}+R_{\theta_A,\theta_B+\pi}+R_{\theta_A+\pi,\theta_B}+R_{\theta_A+\pi,\theta_B+\pi}}
\end{equation}
Finally, the image of $S$ values shown in Fig.~\ref{Figure6}.b is obtained using the following equation:
\begin{equation}
S=|E_{\pi/2,\pi/4}-E_{\pi/2,5\pi/4}|+|E_{0,\pi/4}+E_{0,5 \pi/4}|
\end{equation} 
As shown in Fig.~\ref{Figure6}, $10789$ pairs of $S$ values measured between Alice and Bob correlated pixels show violation of CHSH inequality $S > 2$, over the total of $14129$ pair of pixels forming the two half disks. A spatial averaged value of $\langle S \rangle = 2.20 \pm 0.003 > 2$ is estimated by calculating the mean and variance of $S$ values over these $14129$ pairs of pixels. See the SI section 2 for further details on the CHSH measurement, including Figure B showing all $16$ measured correlation images $R_{\theta_A,\theta_B}$.\\

\noindent \textbf{Acknowledgements.} The Authors acknowledge discussions with M. Barbieri. D.F. acknowledges financial support from the UK Engineering and Physical Sciences Research Council (grants EP/M01326X/1 and EP/R030081/1) and from the European Union's Horizon 2020 research and innovation programme under grant agreement No 801060. H.D. acknowledges funding from the European Union's Horizon 2020 research and innovation programme under the Marie Skłodowska-Curie grant agreement No. 840958.\\
\\
\noindent \textbf{Authors contributions.} H.D. conceived the original idea, designed and performed the experiment, and analysed the data. H.D., A.L., B.N. and D.F. contributed to the interpretation of the results and manuscript. H.D. prepared the manuscript. D.F. supervised the project.\\
\\
\noindent \textbf{Data availability.} Data that support the plots within this paper and other findings of this study are available from DOI:10.5525/gla.researchdata.1093.

\newpage

\clearpage

\part*{Supplementary information}

\section{Details on intensity correlation measurement}

This section provides more details about the intensity correlation measurement performed by the EMCCD camera. Further theoretical details can be found in~\cite{defienne_general_2018-2}.

An EMCCD camera can be used to reconstruct the spatial (a) intensity distribution $I(\vec{k})$ and (b) intensity correlation distribution $\Gamma(\vec{k_1},\vec{k_2})$ of photon pairs, where $\vec{k}$, $\vec{k_1}$ and $\vec{k_2}$ correspond to positions of camera pixels. To do that, the camera first acquires a set of $N$ frames $\{I_l \} _{l \in  [\![ 1,N]\!]}$ using a fixed exposure time. Then: 
\begin{enumerate}[(a)]
\item The intensity distribution is reconstructed by averaging over all the frames:
\begin{equation}
I(\vec{k}) = \frac{1}{N} \sum_{l=1}^N I_l(\vec{k})
\end{equation}
\item The intensity correlation distribution is reconstructed by performing the following substation:
\begin{equation}
\Gamma(\vec{k_1},\vec{k_2}) = \frac{1}{N} \sum_{l=1}^N I_l(\vec{k_1})I_l(\vec{k_2}) - \frac{1}{N-1} \sum_{l=1}^{N-1} I_l(\vec{k_1})I_{l+1}(\vec{k_2})
\end{equation}
Under illumination by the photon pairs, intensity correlations in the left term of the subtraction originate from detections of both real coincidences (two photons from the same entangled pair) and accidental coincidences (two photons from two different entangled pairs), while intensity correlations in the second term originate only from photons from different entangled pairs (accidental coincidence) because there is zero probability for two photons from the same entangled pair to be detected in two succesive images. A subtraction between these two terms leaves only genuine coincidences, that are proportional to the spatial joint probability distribution of the pairs.
\end{enumerate}

\begin{figure*}
\centering
\includegraphics[width=1 \textwidth]{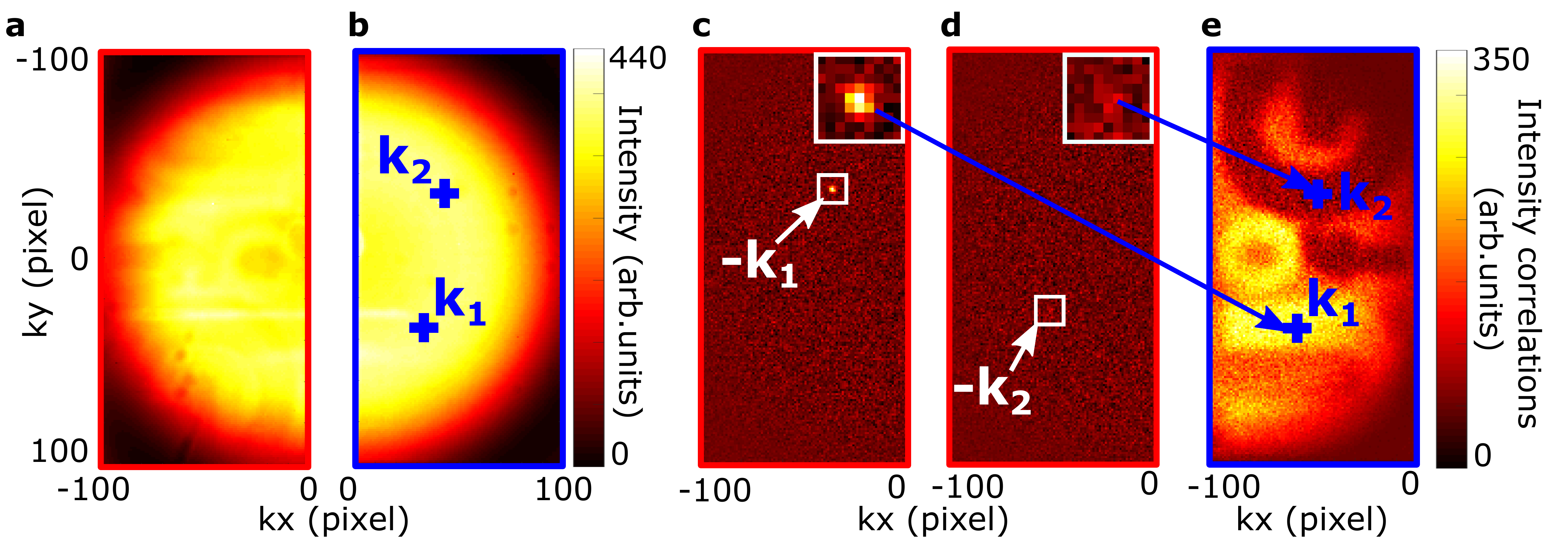} 
\caption{\label{FigureSM0} \textbf{Intensity and intensity correlation measurements between photon pairs with an EMCCD camera.} \textbf{a} and \textbf{b}, Intensity images measured respectively by Alice and Bob. Two pixels $\vec{k_1}$ and $\vec{k_2}$ are arbitrarily selected in Bob side. \textbf{c}, Conditional image relative to position $\vec{k_1}$. \textbf{d}, Conditional image relative to position $\vec{k_2}$. Zooms of areas centred around pixels $\vec{-k_1}$ and $\vec{-k_2}$ are shown in inset. \textbf{e}, Intensity correlation image reconstructed by Bob. Values at $\vec{k_1}$ and $\vec{k_2}$ correspond to intensity correlation values measured at $\vec{-k_1}$ and $\vec{-k_2}$ in images \textbf{c} and \textbf{d}.  }
\end{figure*}

In our work, we use this technique for measuring correlation between pairs of symmetric pixels $\vec{k}$ and $\vec{-k}$ to reconstruct intensity correlation images $R(\vec{k})$. These images correspond exactly to the anti-diagonal component of the complete intensity correlation distribution $R(\vec{k}) = \Gamma(\vec{k},\vec{-k})$. Fig.~\ref{FigureSM0} illustrates these different types of measurements in the case of the experiment described in Fig.1 of the manuscript when Bob displays a phase shift $+\pi$ on his SLM. Fig.~\ref{FigureSM0}.a and b show respectively intensity images measured by Alice (pixels $\vec{k} = (k_x<0,k_y)$) and Bob (pixels $\vec{k} = (k_x>0,k_y)$). These images do not provide any information about the correlation between photon pairs. Fig.~\ref{FigureSM0}.c is the conditional image $\Gamma(\vec{k},\vec{k_1})$ that represents the probability of measuring a photon from a pair at pixel $\vec{k}$ in Alice side conditioned by the detection of its twin photon at pixel $\vec{k_1}$ in Bob side. We observe a strong peak of correlation centred around the symmetric pixel $\vec{-k_1}$ due to the strong anti-correlation between momentum of the pairs (zoom in inset). Similarly, Fig.~\ref{FigureSM0}.d is the conditional image $\Gamma(\vec{k},\vec{k_2})$ relative to position $\vec{k_2}$. In this last case, the peak of correlation is very weak (zoom in inset). Finally, Fig.~\ref{FigureSM0}.e shows the intensity correlation image $R(\vec{k}) = \Gamma(\vec{k},\vec{-k})$. In this last image, the value at pixel $\vec{k_1}$ corresponds to the value of the peak of correlation at $\vec{-k_1}$ shown in Fig.~\ref{FigureSM0}.c and the value at pixel $\vec{k_2}$ corresponds to the value of the peak of correlation at $\vec{-k_2}$ in Fig.~\ref{FigureSM0}.d.

\section{Details on spatial and polarisation entanglement of the source}
\label{sec1}
As described in Fig.3 of the manuscript, entangled photon pairs are generated by type I SPDC using a pair of BBO crystals. These pairs are entangled in both their polarisation and spatial degree of freedom~\cite{barreiro_generation_2005}.

\subsection{Spatial entanglement}

Spatial entanglement between photons is characterised by performing intensity correlation measurement between (a) positions and (b) momentum of photons~\cite{howell_realization_2004,moreau_realization_2012,tasca_imaging_2012}. Fig.~\ref{FigureSM6} describes the corresponding experimental apparatus:
\begin{enumerate}[(a)]
\item Positions $\vec{r}$ of photons are mapped onto pixels of the camera using a two-lens imaging configuration $f_1-f_2$. After measuring the intensity correlation distribution $\Gamma(\vec{r_1},\vec{r_2})$, its projection along the minus-coordinate axis $\vec{r_1-\vec{r_2}}$ is shown in Fig.~\ref{FigureSM6}.b. The peak of correlation at its center is a signature of strong correlations between positions of photons. The position correlation width in the camera plane $\sigma_r^{(c)} = 26.20 \pm 0.02 \upmu$m $\approx 1.6$ pixel is estimated by fitting the minus-coordinate projection by a Gaussian model~\cite{fedorov_gaussian_2009} of the form $a \exp(-|\vec{k_1}+\vec{k_2}|^2/(2\sigma_k^{(c)^2})$. To obtain a better estimate of the correlation width taking into account the approximation made by using the  Gaussian model, a correction factor $\beta = \sqrt{\alpha/(\alpha+\alpha^{-1})}$ with $\alpha = 0.455$~\cite{chan_transverse_2007,schneeloch_introduction_2016} is applied to the measured width resulting into $\sigma_r = \beta \sigma_r^{(c)} = 10.85 \pm 0.06$ $\upmu$m. Note that the use of such correction procedure is not crucial in our work because an overestimated value of the product $\sigma_k \sigma_r^{(c)} = [3.71 \pm 0.01].10^{-2} < \frac{1}{2}$ still violates significantly the EPR-type inequality~\cite{howell_realization_2004}.

\item Momentum $\vec{k}$ of photons are mapped onto pixels of the camera by replacing the lens $f_2$ by a lens with twice its focal length. After measuring the intensity correlation distribution $\Gamma(\vec{k_1},\vec{k_2})$, its projection along the sum-coordinate axis $\vec{k_1+\vec{k_2}}$ is shown in Fig.~\ref{FigureSM6}.c. The peak of correlation at its center is a signature of strong anti-correlations between momentum of photons. The momentum correlation width in the camera plane $\sigma_k^{(c)} = 17 \pm 0.01 \upmu$m $\approx 1.1$ pixel is estimated by fitting the sum-coordinate projection by a Gaussian model~\cite{fedorov_gaussian_2009} of the form $a \exp(-|\vec{k_1}+\vec{k_2}|^2/(2\sigma_k^{(c)^2})$. The corresponding value in the momentum space $\sigma_k$ is then calculated using the formula $\sigma_k = 2 \pi \sigma_k^{(c)} /(\lambda f_2) = [1.326 \pm 0.001]\times10^3\,$rad.m$^{-1}$ where $f_2 = 100$mm and $\lambda=810$nm.
\end{enumerate}

The presence of spatial entanglement between photons is certified by violating an EPR-type inequality: $ \sigma_r \sigma_k = [1.44 \pm 0.01]\times10^{-2} < \frac{1}{2}$~\cite{howell_realization_2004}. 

\subsection{Polarisation entanglement}

\begin{figure*}
\centering
\includegraphics[width=1 \textwidth]{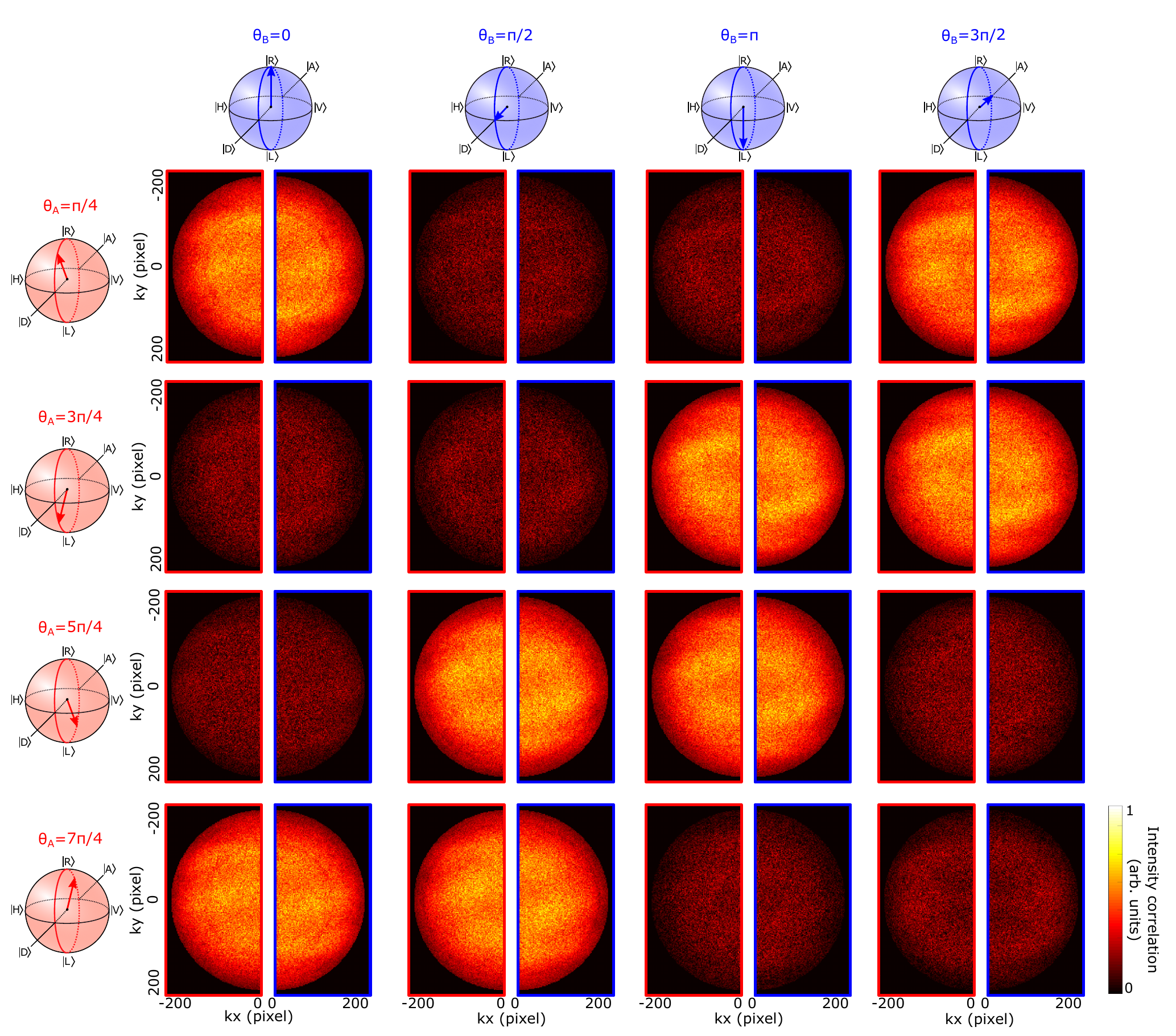} 
\caption{\label{FigureSM16} \textbf{Intensity correlation images measured by Alice and Bob for $16$ combinations of phase values $\theta_A$ and $\theta_B$.} Intensity correlation images are shown by pair, with a red outline for Alice and a blue outline for Bob. Each row corresponds to a measurement setting on Alice SLM $\theta_A=\{\pi/4,3 \pi/4,5\pi/4,7\pi/4 \}$ and each column to a measurement setting of Bob SLM $\theta_B=\{0,\pi/2,\pi,3 \pi/2 \}$.  }   
\end{figure*}

\begin{figure*}
\centering
\includegraphics[width=0.85 \textwidth]{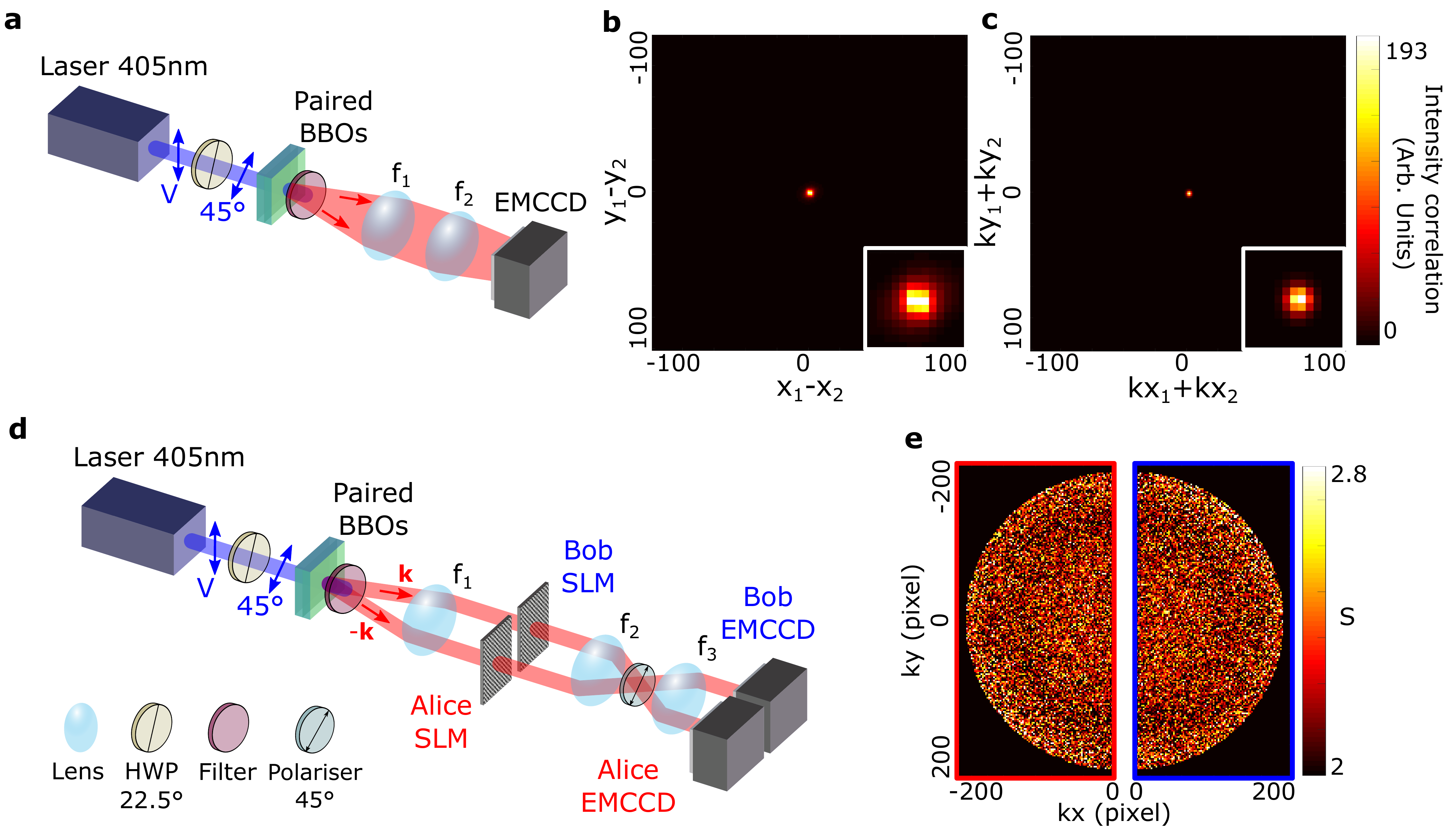} 
\caption{\label{FigureSM6} \textbf{Spatial and polarisation entanglement characterisation.} \textbf{a}, Experimental setup used for spatial entanglement characterisation. $f_1=75$mm and $f_2=100$mm form a two-lenses imaging system. \textbf{b}, Minus-coordinate projection of the intensity correlation distribution measured between positions of photons. Zoom of the central area is in inset. The width of the peak $\sigma_r^{(c)} = 26.20 \pm 0.02 \upmu$m is measured using a Gaussian model of the form $a \exp(-|\vec{x_1}-\vec{x_2}|/(2 \sigma_r^{(c)^2}))$. \textbf{c}, Sum-coordinate projection of the intensity correlation distribution measured between momentum of photons. In this case, the lens $f_2$ is replaced by a lens of twice smaller focal length $f_2'=50$mm to map momentum of photons onto pixels of the camera. Zoom of the central area is in inset. The width of the peak $\sigma_k^{(c)} = 17.00 \pm 0.01 \upmu$m is measured using a Gaussian model of the form $a \exp(-|\vec{k_1}+\vec{k_2}|/(2 \sigma_r^{(c)^2}))$. \textbf{d}, Experimental setup used for polarisation entanglement characterisation. $f_1=54$mm is positioned in a Fourier imaging configuration and $f_2=150$ and $f_3=100$mm for a two-lenses imaging configuration. \textbf{e}, Image of measured Clausen-Horne-Shimony-Holt (CHSH) inequality values, denoted $S$.}
\end{figure*}

The presence of polarisation entanglement between photons can be demonstrated by violating the Clauser-Horne-Shimony-Holt (CHSH) inequality~\cite{clauser_proposed_1969}. A simplified version of the experimental apparatus used to perform this measurement is shown in Fig.~\ref{FigureSM6}.d. In this case, the combination of Alice and Bob SLMs with a $45$ degrees polariser positioned in front of the cameras play the role of the rotating polarisers used in a conventional CHSH violation experiment~\cite{aspect_experimental_1982}. When a constant phase $\theta_B$ ($\theta_A$) is programmed onto Bob SLM (Alice SLM), each pixel of Bob camera (Alice) performs a measurement that corresponds to an operator of the form:
\begin{equation}
\hat{B}_{\theta_B} = \frac{1}{2} \big[ \ket{H}  \bra{H}  + \ket{V}  \bra{V} + e^{i \theta_B} \ket{H}  \bra{V}  + e^{-i \theta_B} \ket{V}  \bra{H} \big]
\end{equation} 
In the first step of this experiment, Alice and Bob measure intensity correlation images $R_{\theta_A,\theta_B}$ for $16$ combinations of phase values $\theta_A$ and $\theta_B$ programmed on their SLMs. On the one hand, Bob uses the same set as the one used in the holographic process $\theta_B=\{0,\pi/2,\pi,3 \pi/2 \}$, where $\theta_B=0$ and $\theta_B=\pi$ correspond to measurements performed in the diagonal polarisation basis $ \hat{B}^{\pm}_{0/\pi} = (\ket{V} \pm \ket{H}) (\bra{V} \pm \bra{H})$ and values $\theta_B=\pi/2$ and $\theta_B=3 \pi/2$ correspond to measurements performed in the circular polarisation basis $ \hat{B}^{\pm}_{\frac{\pi}{2}/\frac{3\pi}{2}} = (\ket{V} \pm i \ket{H}) (\bra{V} \mp i \bra{H})$. On the other hand, Alice programs phase values $\theta_A=\{\pi/4,3 \pi/4,5\pi/4,7\pi/4 \}$ of her SLM to perform measurements in diagonal and circular basis rotated by $\pi/4$: $\hat{A}^{\pm}_{\frac{\pi}{4}/\frac{5\pi}{4}} = (\ket{V} \pm \frac{1}{\sqrt{2}} \ket{H}) (\bra{V} \pm \frac{1}{\sqrt{2}} \bra{H})$ and $ \hat{A}^{\pm}_{\frac{3\pi}{4}/\frac{7\pi}{4}} = (\ket{V} \pm \frac{i}{\sqrt{2}} \ket{H}) (\bra{V} \mp \frac{i }{\sqrt{2}} \bra{H})$. Coincidence images associated with these $4\times4=16$ measurement combinations are shown in Figure~\ref{FigureSM16}. In the second step, Alice and Bob combine these $16$ intensity correlation images to compute  $E_{\theta_A, \theta_B}$ using the following formula~\cite{clauser_proposed_1969}:
\begin{equation}
E_{\theta_A,\theta_B} = \frac{R_{\theta_A,\theta_B}-R_{\theta_A,\theta_B+\pi}-R_{\theta_A+\pi,\theta_B}+R_{\theta_A+\pi,\theta_B+\pi}}{R_{\theta_A,\theta_B}+R_{\theta_A,\theta_B+\pi}+R_{\theta_A+\pi,\theta_B}+R_{\theta_A+\pi,\theta_B+\pi}}
\end{equation}
Finally, an image of $S$ values is obtained using the following equation:
\begin{equation}
S=|E_{\pi/2,\pi/4}-E_{\pi/2,5\pi/4}|+|E_{0,\pi/4}+E_{0,5 \pi/4}|
\end{equation} 
This image is shown in Fig.5.b of the manuscript and in Fig.~\ref{FigureSM6}.e. We observe that $10789$ values of $S$ measured between in Alice and Bob correlated pixels show violation of the CHSH inequality $S > 2$, over the total of $14129$ within the two half-disks. A spatial averaged value of $\langle S \rangle = 2.20 \pm 0.003 > 2$ is estimated by calculating the mean of $S$ values over the two half-disks areas, and the error on $\langle S \rangle$ is calculated from their variance. Violation of CHSH inequality demonstrates the presence of polarisation entanglement between photon pairs.

\section{Details on signal-to-noise and spatial resolution}

This section provides more details about the signal-to-noise ratio (SNR) and spatial resolution of the phase image retrieved by quantum holography.

\subsection{Signal-to-noise ratio.}

In our work, we define the SNR as the ratio between $\pi$ and the standard deviation of the noise measured in a region of the image in which phase values equal $\pi$. It is for example the case for the area within the letters \textit{U} and \textit{o} in the phase image shown in Fig.1.b. For a constant source intensity and a fixed exposure time, the factor that most influences the SNR is the total number of images $N$ acquired to measure the four intensity correlation images used to reconstruct the phase image. Each phase image shown in the manuscript has been retrieved using $N=2.5\times10^6$ images and show SNR values ranging between $19$ and $21$. Fig.\ref{FigureSM1}.a shows SNR values measured for different values of $N$ (black crosses). As predicted by the theory~\cite{defienne_general_2018-2} and demonstrated by fitting the data (blue dashed curve), the SNR evolves as $\sqrt{N}$. Fig.\ref{FigureSM1}.b-c show three images of the retrieved phase for respectively $N=2.5\times10^4$ images, $N=2.5\times10^5$ and $N=2.5\times10^7$ images.

\subsection{Spatial resolution.} 
\begin{figure*}
\centering
\includegraphics[width=0.8 \textwidth]{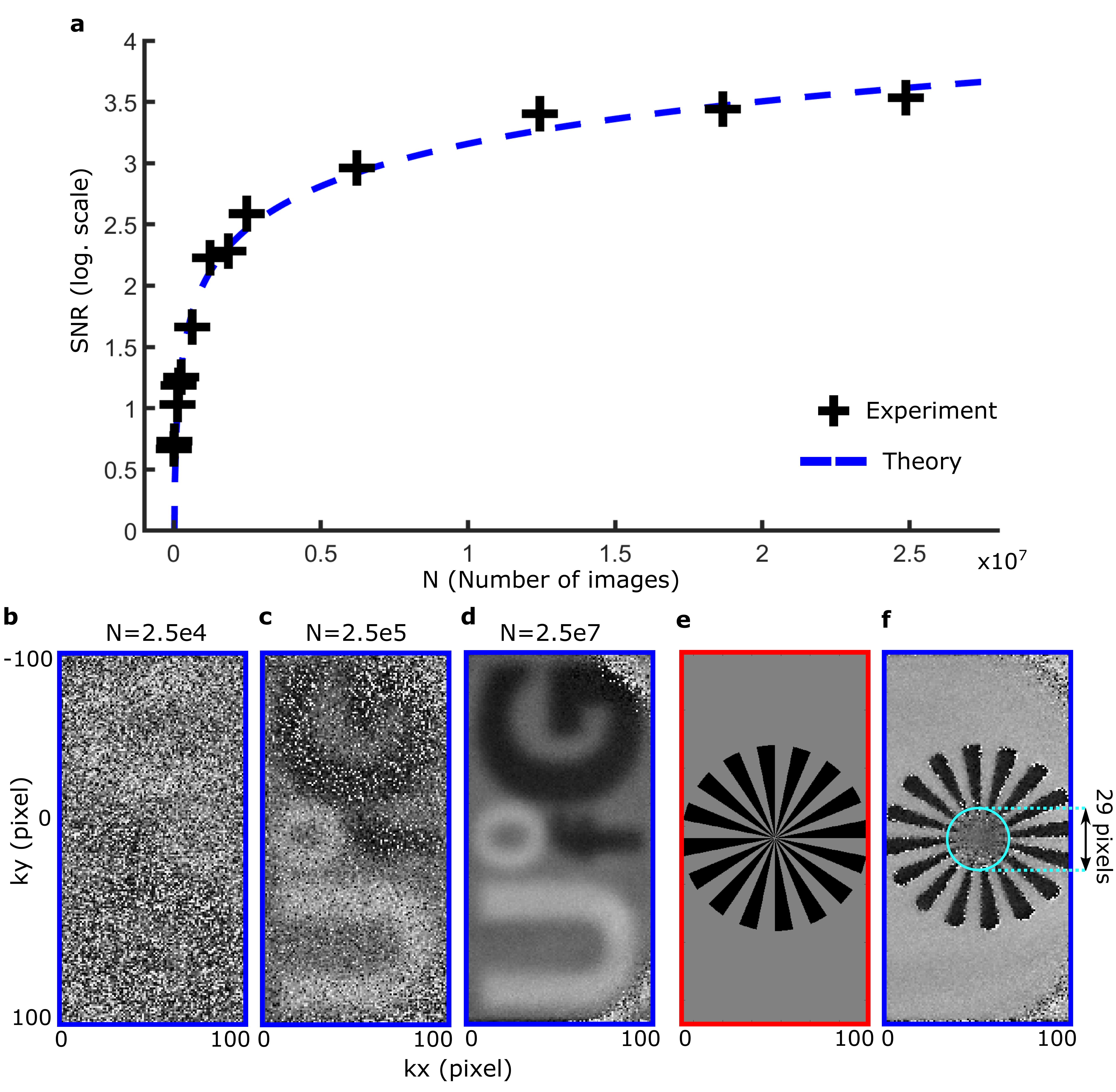} 
\caption{\label{FigureSM1} \textbf{Signal-to-noise and spatial resolution}. \textbf{a}, Values of single-to-noise ratio (SNR) in the retrieved phase image measured for different total number of images acquired $N$ (black crosses) together with a theoretical fit of the form: SNR$= 0.074 \sqrt{N}$ (blue dashed line). \textbf{b}, Phase image reconstructed using $N=2.5\times10^4$ images. \textbf{b}, Phase image reconstructed using $N=2.5\times10^5$images. \textbf{b}, Phase image reconstructed using $N=2.5\times10^7$images. \textbf{e}, Radial resolution target (Siemens star with 16 branches) phase pattern programmed on Alice SLM. \textbf{f}, Reconstructed phase image of the radial resolution target. Resulting spatial resolution is $d = 45 \pm 3$ $\upmu$m, which corresponds to approximately to $2.8$ pixels }
\end{figure*}

To measure the spatial resolution in the retrieved phase image, a radial resolution target (Siemens star with 16 branches) is programmed by Alice Fig.~\ref{FigureSM1}.e. Fig.~\ref{FigureSM1}.f shows the retrieved image. When comparing this retrieved image (Fig.~\ref{FigureSM1}.f) to the ground truth (Fig.~\ref{FigureSM1}.e), the area of the resolution target that is not spatially resolved is approximately a disk of diameter $29$ pixels, which corresponds to a perimeter of $91$ pixels. The spatial resolution is then estimated by dividing the perimeter by twice the number of branches $91/(2*16) \approx 2.8$ pixels, which results to a spatial resolution of $d = 45 \pm 3$ $\upmu$m. We observe that this value of the spatial resolution is larger than this of the momentum correlation width ($\sigma_k = 18 \upmu$m on the camera, see Fig.~\ref{FigureSM6}) which in theory should be the resolution  limit of our system. The origin of this difference is that the SLM on which phase objects are programmed is not perfectly positioned into an image plane of the system, which is a misalignment that can be expected when using an SLM acting in reflection and tilted. This misalignment causes a small blurring of the phase object programmed on the SLM and thus lead to an overestimation of the measured value of the spatial resolution width.  

\section{Details on phase distortion characterisation}

This section provides more details about the characterisation of the phase distortion $\Psi_0$.

\begin{figure*}
\centering
\includegraphics[width=0.8 \textwidth]{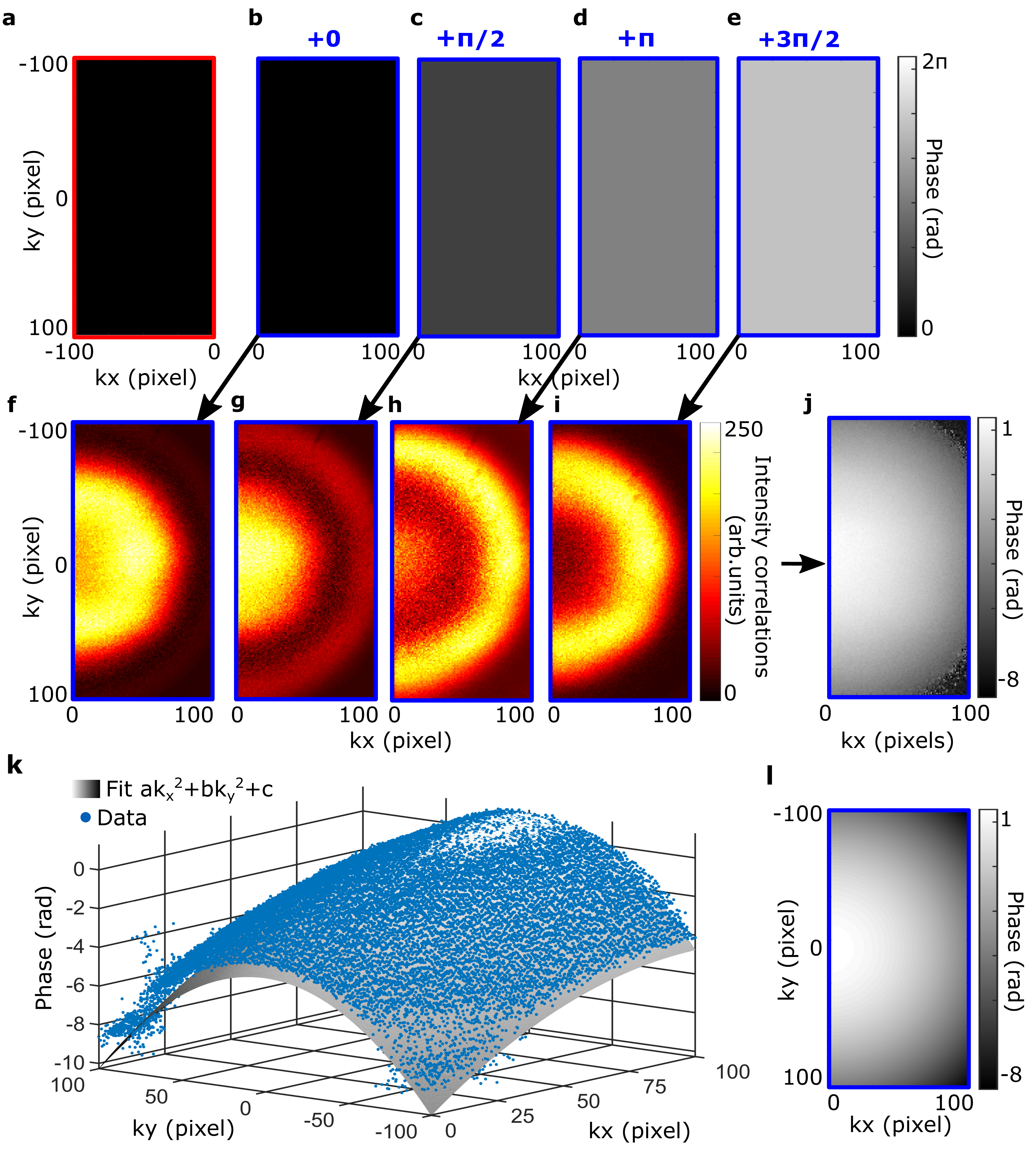} 
\caption{\label{FigureSM4} \textbf{Phase distortion characterisation.} \textbf{a}, Flat phase pattern programmed on Alice SLM. \textbf{b-e}, Phase-shifted patterns programmed on Bob SLM: $+0$ (\textbf{b}), $+\pi/2$ (\textbf{c}), $+\pi$ (\textbf{d}) and $+3 \pi/2$ (\textbf{e}). \textbf{f-i}, Intensity correlation images measured for each phase mask displayed on Bob SLM. \textbf{j}, Phase image retrieved by Bob. \textbf{k}, Fit of the phase image by a quadratic function of the form: $\Psi_0(k_x,k_y) = 4.69 k_x^2+5.04 k_y^2+0.02$. \textbf{l}, Correction phase pattern resulting from the fitting process.}
\end{figure*}

Phase distortion $\Psi(\vec{k})$ originates from the SPDC process used to produce pairs of photons~\cite{hegazy_tunable_2015,hegazy_relative-phase_2017}. To characterise it, Alice and Bob perform a quantum holographic experiment using the scheme described in Fig.1.a. In this case, Alice programs a flat phase pattern $\theta_A=0$ (Fig.~\ref{FigureSM4}.a) and Bob programs successive phase shifts patterns with values $+0$ (Fig.~\ref{FigureSM4}.b), $+\pi/2$ (Fig.~\ref{FigureSM4}.b), $+\pi$ (Fig.~\ref{FigureSM4}.b) and $+3 \pi/2$ (Fig.~\ref{FigureSM4}.b), without any correction phase mask superimposed on them. Fig.~\ref{FigureSM4}.f-i show intensity correlations images measured for each of the four SLM patterns displayed by Bob. Therefore, the resulting reconstructed phase image corresponds exactly to the phase distortion $\Psi_0(\vec{k})$ (Fig.~\ref{FigureSM4}.j). The phase image is then fitted by a quadratic function of the form $\Psi_0(k_x,k_y) = 4.69 k_x^2+5.04 k_y^2+0.02$ (Fig.~\ref{FigureSM4}.k). Result of the fit is then used to construct and program the phase compensation pattern on Bob SLM (Fig.~\ref{FigureSM4}.l).  

\section{Details on quantum holography without polarisation entanglement }

This section provides more details about quantum holography performed with photons entangled in space but not in polarisation.

\begin{figure*}
\centering
\includegraphics[width=1\textwidth]{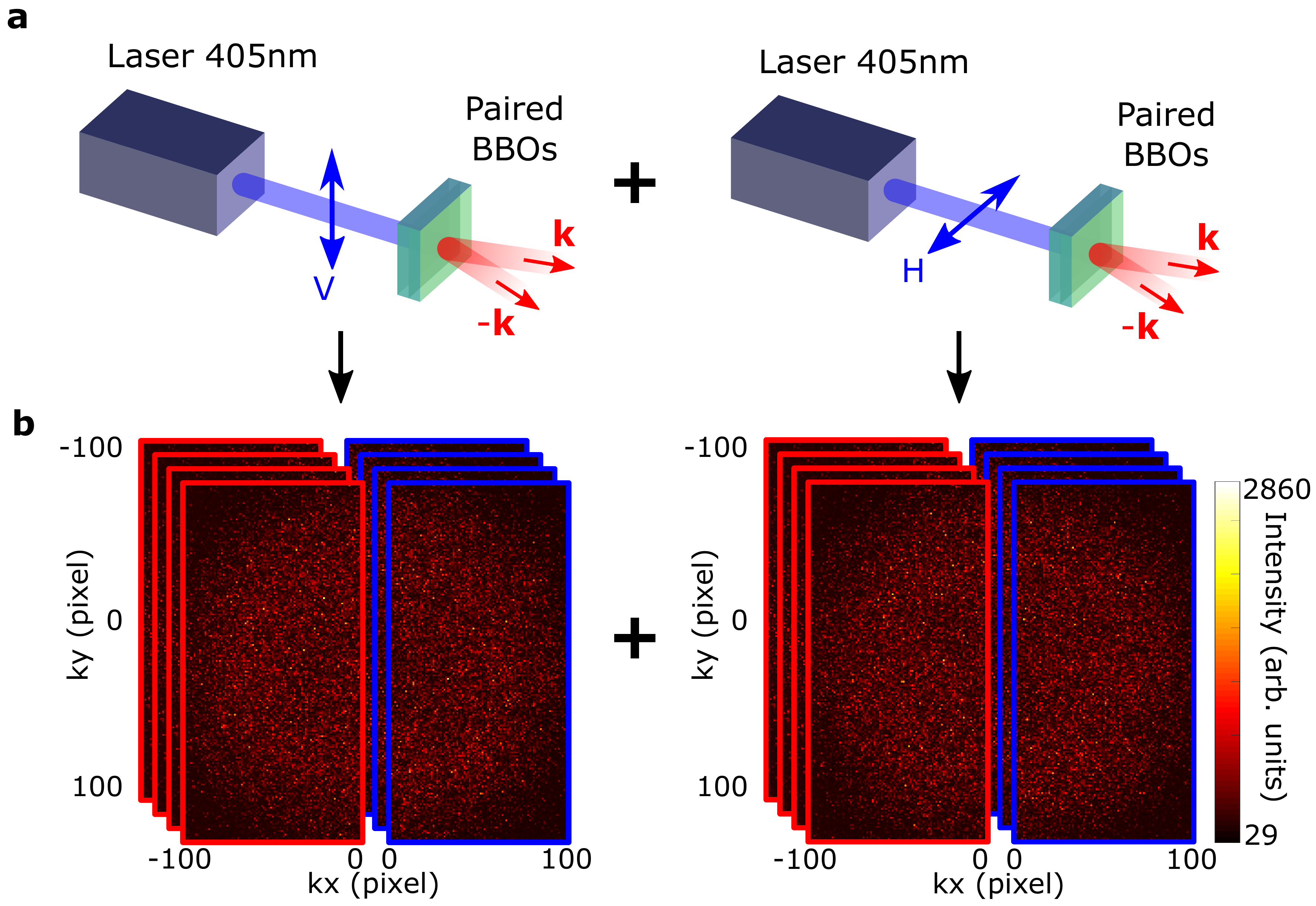} 
\caption{\label{FigureSM5} \textbf{Generation of a mixed state without polarisation entanglement.} \textbf{a}, Experimental configurations used to generate the mixed state: $\sum_{\vec{k}} \big[ \ket{H}_\vec{\vec{k}}  \ket{H}_{-\vec{k}} \bra{H}_\vec{\vec{k}}  \bra{H}_{-\vec{k}}  + \ket{V}_\vec{k}  \ket{V}_{-\vec{k}} \bra{V}_\vec{k}  \bra{V}_{-\vec{k}} \big]$. Polarisation of pump laser is alternated between vertical and horizontal. \textbf{b}, Frames acquired in each configuration are summed. Intensity images shown in Fig.2.b.c of the manuscript and intensity correlation images shown in Fig.2.e-f of the manuscript are reconstructed from the set of summed frames.   }
\end{figure*}

Fig.2 of the manuscript shows results of quantum holographic experiment performed with photons that are entangled in space but not in polarisation. In this experiment, the state generated at the output of the crystals is defined by the density operators:
\begin{equation}
\frac{1}{2}\sum_{\vec{k}} \big[ \ket{H}_\vec{\vec{k}}  \ket{H}_{-\vec{k}} \bra{H}_\vec{\vec{k}} \bra{H}_{-\vec{k}} + \ket{V}_\vec{k}  \ket{V}_{-\vec{k}} \bra{V}_\vec{k}  \bra{V}_{-\vec{k}} \big]
\end{equation}
This state is composed of a balanced statistical mixture of two pure states: $\sum_{\vec{k}} \ket{H}_\vec{\vec{k}}  \ket{H}_{-\vec{k}}$ and $\sum_{\vec{k}} \ket{V}_\vec{\vec{k}}  \ket{V}_{-\vec{k}}$. As shown in Fig.~\ref{FigureSM5}.a, this state is produced by alternating the polarisation of the pump laser between vertical and horizontal polarisations. Frames acquired by the camera in each configuration are then summed together (Fig.~\ref{FigureSM5}.b). Then, the intensity and intensity correlation images shown in Fig.2 of the manuscript are reconstructed using the set of summed frames. This experiment is equivalent of generating the photons with a pump that is coherent in space and time but not in polarisation (i.e. unpolarised). Due to the fundamental transfer of coherence that occurs between the pump and the down-converted two-photon field in SPDC~\cite{jha_spatial_2010,kulkarni_intrinsic_2016,kulkarni_transfer_2017}, the produced state is entangled in space and time, but not in polarisation.

Note that, even if entanglement is absent, we still observe some small spatial inhomogeneities in the reconstructed phase in the Figure.2.d of the manuscript. These inhomogeneities originate from the small spatial variations observed in the corresponding intensity correlation images (Figs.2.f-i of the manuscript), that themselves originate from the small and not desired intensity modulations performed by the SLM during the phase shifting process. This effect is also present in the measurements performed with polarisation entanglement, but it is not visible in the reconstructed phase (Fig.1.j) because the corresponding spatial variations in intensity correlations images (Figs.1.f-i) are negligible compared to those created by the object.

\section{Details on the characterisation of the dynamic phase disorder}

This section provides more details about the dynamic phase disorder and its characterisation. 

\begin{figure*}
\centering
\includegraphics[width=1 \textwidth]{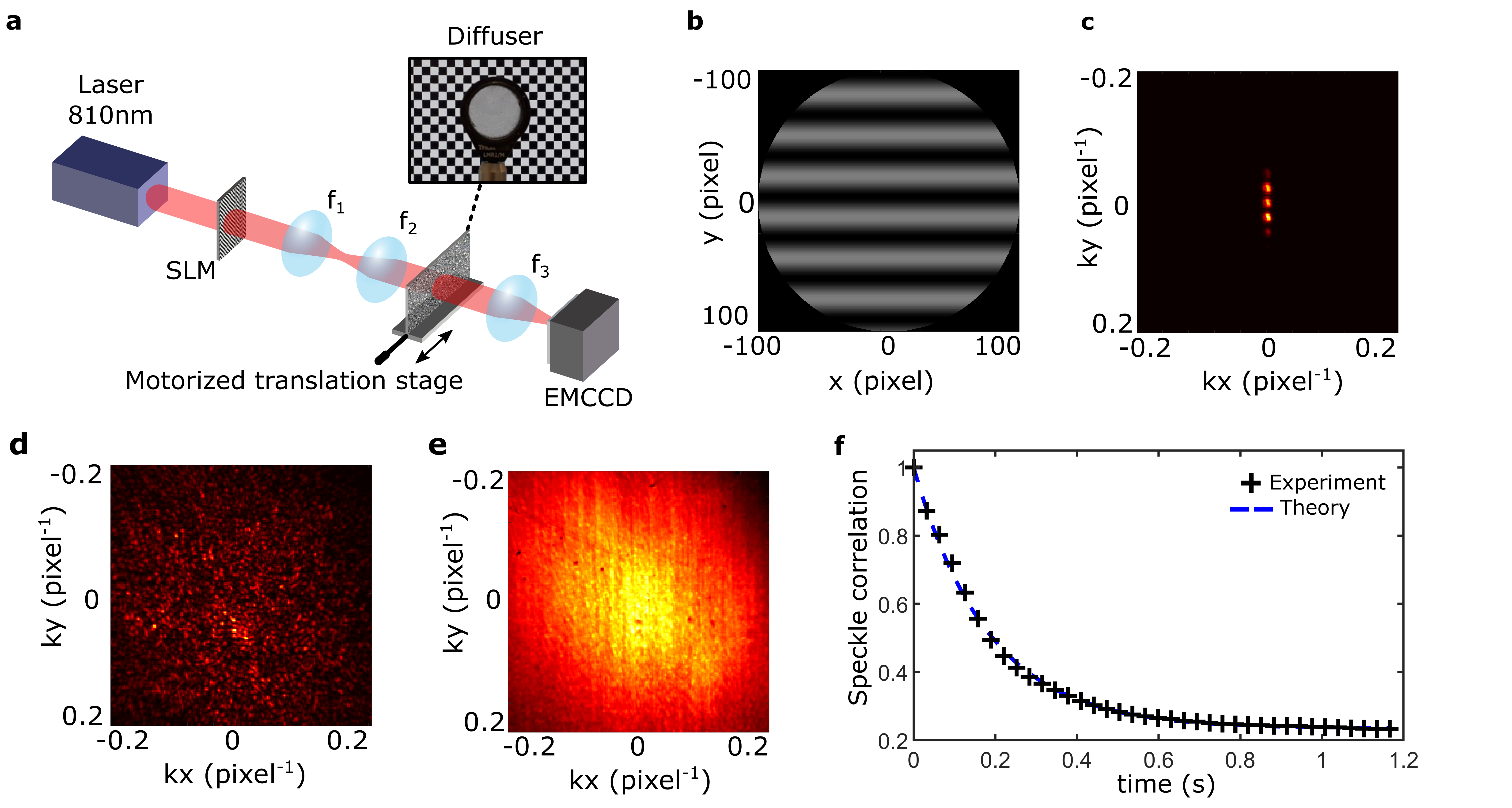} 
\caption{\label{FigureSM2} \textbf{Characterisation of the dynamic phase disorder}. \textbf{a}, A classical laser ($810$nm ; beam diameter $0.8$mm) illuminates an SLM that is imaged onto a the diffuser by two lenses $f_1=150$mm and $f_2=75$mm. Lens $f_3=75$mm Fourier-images the diffuser onto the camera. \textbf{b}, Sine-shaped phase pattern programmed on the SLM. \textbf{c}, Intensity image measured on the camera without diffuser. \textbf{d}, Intensity image measured on the camera with the diffuser maintained static. \textbf{e}, Intensity image measured on the camera with the moving diffuser using an exposure time of $0.5$s. Width of the diffuse halo is estimated to $1.3$mm by fitting with a Gaussian function. \textbf{f}, Measurement of speckle correlation values with time (black crosses) together with a theoretical fit of the form $1-0.77 (1-e^{-t/0.183})$ (blue dashed line). Typical decorrelation time equals to $183$ms.     }
\end{figure*}

Fig.~\ref{FigureSM2}.a shows the experimental setup used to characterise properties of the phase disorder introduce by the presence of the diffuser, that is a plastic sleeve layer of thickness $<100$ $\upmu$m.  Without diffuser, a sine-shaped phase image programmed on the SLM (Fig.~\ref{FigureSM2}.b) generates a very specific diffraction pattern on the camera (Fig.~\ref{FigureSM2}.c). After introducing the static diffuser, the diffraction pattern becomes a speckle pattern (Fig.~\ref{FigureSM2}.c). The presence of the diffuser erases all information about the phase image, that cannot be retrieved by classical holographic techniques such as phase retrieval~\cite{fienup_phase_1982} and phase-stepping holography~\cite{yamaguchi_phase-shifting_1997}. When the diffuser is moving, the speckle pattern takes the form of a diffuse halo if the exposure time of the camera is larger ($0.5$s) than the typical decorrelation time of the speckle. The halo width is estimated to $1.3$mm by Gaussian fitting which corresponds approximately to a diffusing angle of $1$ degree and a surface roughness of $46$ $\upmu$m. Moreover, the dynamic properties of the disorder are estimated by measuring the speckle decorrelation time. Speckle correlation coefficients are calculated by acquiring a series of speckle patterns using short exposure times ($3$ms) and correlating them with a reference speckle image. Fig.~\ref{FigureSM2} shows the decrease of speckle correlation with time (black crosses). A typical decorrelation time of $183$ms is measured by fitting data with an exponential model (dashed blue curve). 

\section{Details about the quantum holographic measurements performed in the presence of static and dynamic stray light}

This section provides more details about the quantum holographic measurements performed in the presence of stray light. 

\begin{figure*}
\centering
\includegraphics[width=1 \textwidth]{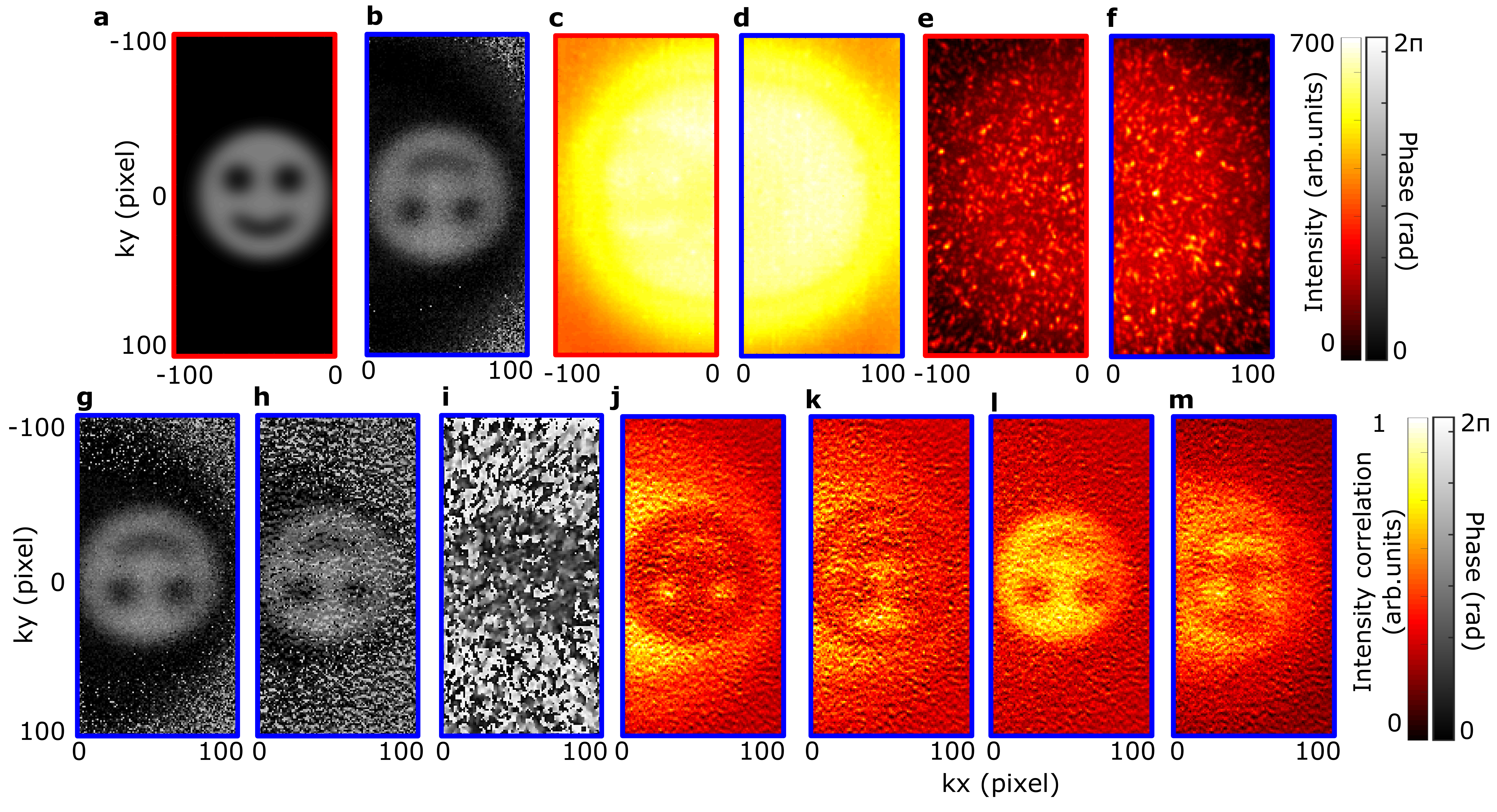} 
\caption{\label{FigureSM11bis} \textbf{Phase reconstruction through dynamic stray light.} \textbf{a}, Phase image programmed by Alice. \textbf{b}, Phase image reconstructed by Bob without stray light (SNR=$12$ and NMSE=$6 \%$). \textbf{c} and \textbf{d}, Intensity images measured by Alice and Bob in the presence of dynamic stray light, respectively. \textbf{e} and \textbf{f}, Intensity images measured by Alice and Bob in the presence of the same stray light but maintained static on purpose to show the complex spatial structure of the time-varying speckle pattern. \textbf{g} and \textbf{h}, Phase images retrieved by Bob with dynamic stray light of average intensity $0.5$ that of quantum light  (SNR=$9$ and NMSE=$17 \%$) and dynamic stray light with intensity $1.2$ that of quantum light (SNR=$3$ and NMSE=$43 \%$), respectively. \textbf{i}, Phase image reconstructed using an equivalent classical holographic system under similar dynamic stray illumination conditions with $0.5$ average intensity ratio (NMSE=$82\%$) (see Figure~\ref{FigureSM8}.b)  \textbf{j-m}, Intensity correlation images measured for each phase shifting mask displayed on Bob SLM, with no traces of classical stray light (intensity ratio $0.5$). All images were reconstructed from $5\times10^6$ frames. }
\end{figure*}

\begin{figure*}
\centering
\includegraphics[width=0.8 \textwidth]{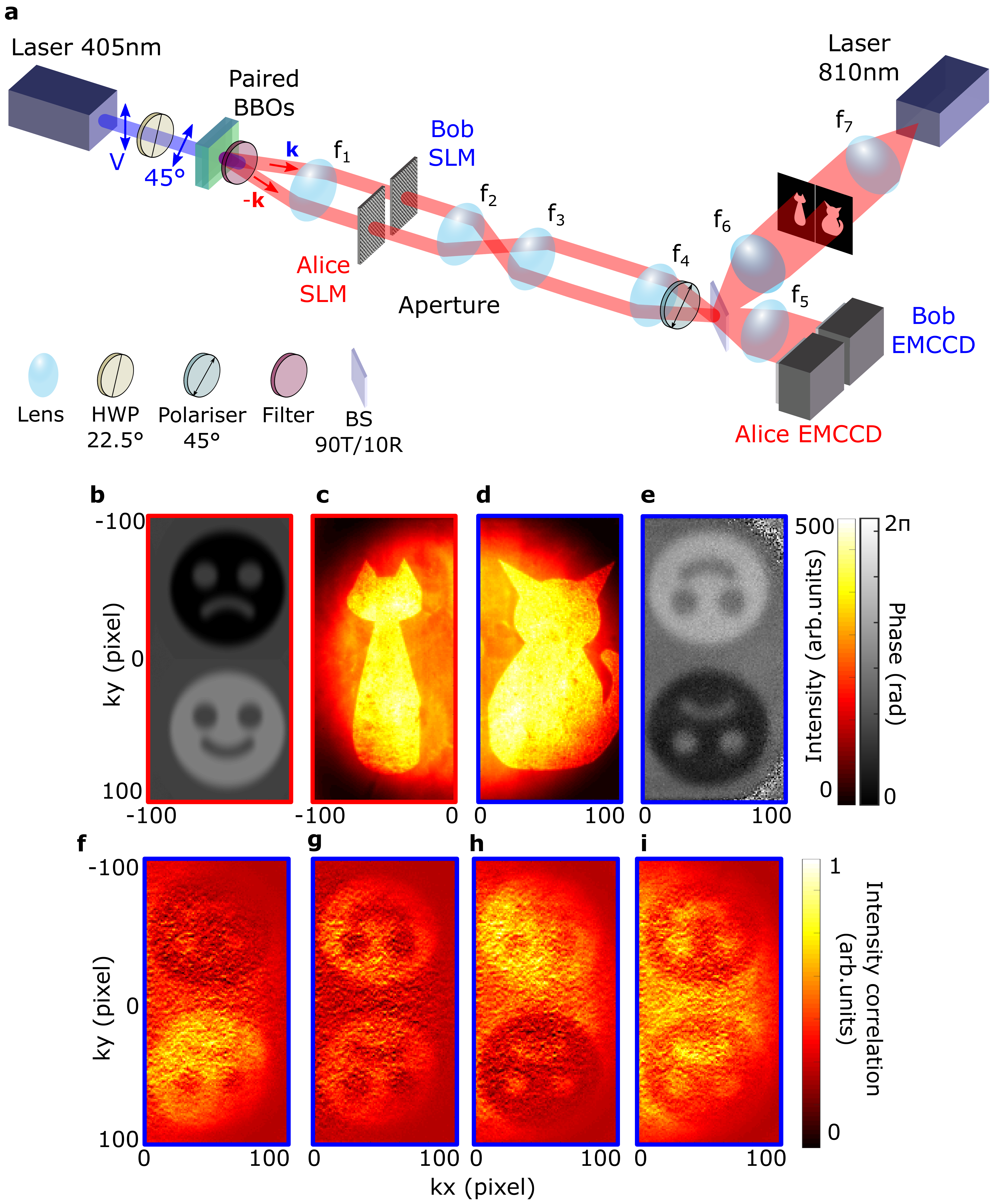} 
\caption{\label{FigureSM11} \textbf{Phase reconstruction through static stray light.} Experimental setup including the source of static stray light i.e. two cat-shaped objects illuminated by a classical laser. \textbf{b}, Phase image encoded by Alice. \textbf{c} and \textbf{d}, Intensity images measured by Alice and Bob with well visible cat-shaped images. \textbf{e}, Phase image retrieved by Bob with SNR$=20$ and NMSE=$5\%$. \textbf{f-i}, Intensity correlation images measured for each phase mask displayed on Bob SLM, with no traces of cat-shaped classical images. The phase image was reconstructed from $5.10^6$ frames.}
\end{figure*}

Figure~\ref{FigureSM11bis} and~\ref{FigureSM11} show experimental results of quantum holographic measurements performed with dynamic and static stray light falling on the sensor, respectively. Figures~\ref{FigureSM11bis}.a,c,d and g are respectively the same images than those shown in Figure 4.e-h of the manuscript. In addition, Figure~\ref{FigureSM11bis}.e and f show intensity images acquired by Alice and Bob in the presence of the same stray light but maintained static on purpose to show the complex spatial shape of the time-varying speckle pattern. Figure~\ref{FigureSM11bis}.b show the phase image reconstructed without stray light (SNR=$12$ and NMSE=$6 \%$) and Figures~\ref{FigureSM11bis}.h shows the phase image reconstructed in the presence of dynamic stray light with intensity ratio quantum/classical of $1.2$. Furthermore, Figures~\ref{FigureSM11}.i-m show the four intensity correlation images used for reconstructing the phase image in the case of the presence of dynamic stray light with ratio $0.5$. No traces of classical stray light appear on any the intensity correlation images~\cite{defienne_quantum_2019}, while they are well visible on the direct intensity images (Figs.~\ref{FigureSM11bis}.c.d). For comparison, Fig.~\ref{FigureSM11bis}i shows a very degraded phase image (NMSE=$82 \%$) retrieved using an equivalent classical holographic system (see Figure~\ref{FigureSM8}.b), under the same dynamic stray light conditions with intensity ratio of $0.5$ (see Methods). To confirm that the absence of classical traces is not due to a spatial averaging effect by the time-varying speckle, we performed the same experiment with static stray light. Figure~\ref{FigureSM11}a shows a scheme of the experimental setup including the static stray light i.e. two cat-shaped objects illuminated by a classical laser. Figures~\ref{FigureSM11}.b-e show the reconstruction of a phase image in the presence of two cat-shaped images superimposed on Alice and Bob sensors. Figures~\ref{FigureSM11}.f-i confirm the absence of classical traces on the intensity correlation images used to reconstruct the phase. This additional experiment confirms that our quantum holographic protocol is insensitive to the dynamics of stray light.

\begin{figure*}
\centering
\includegraphics[width=0.7 \textwidth]{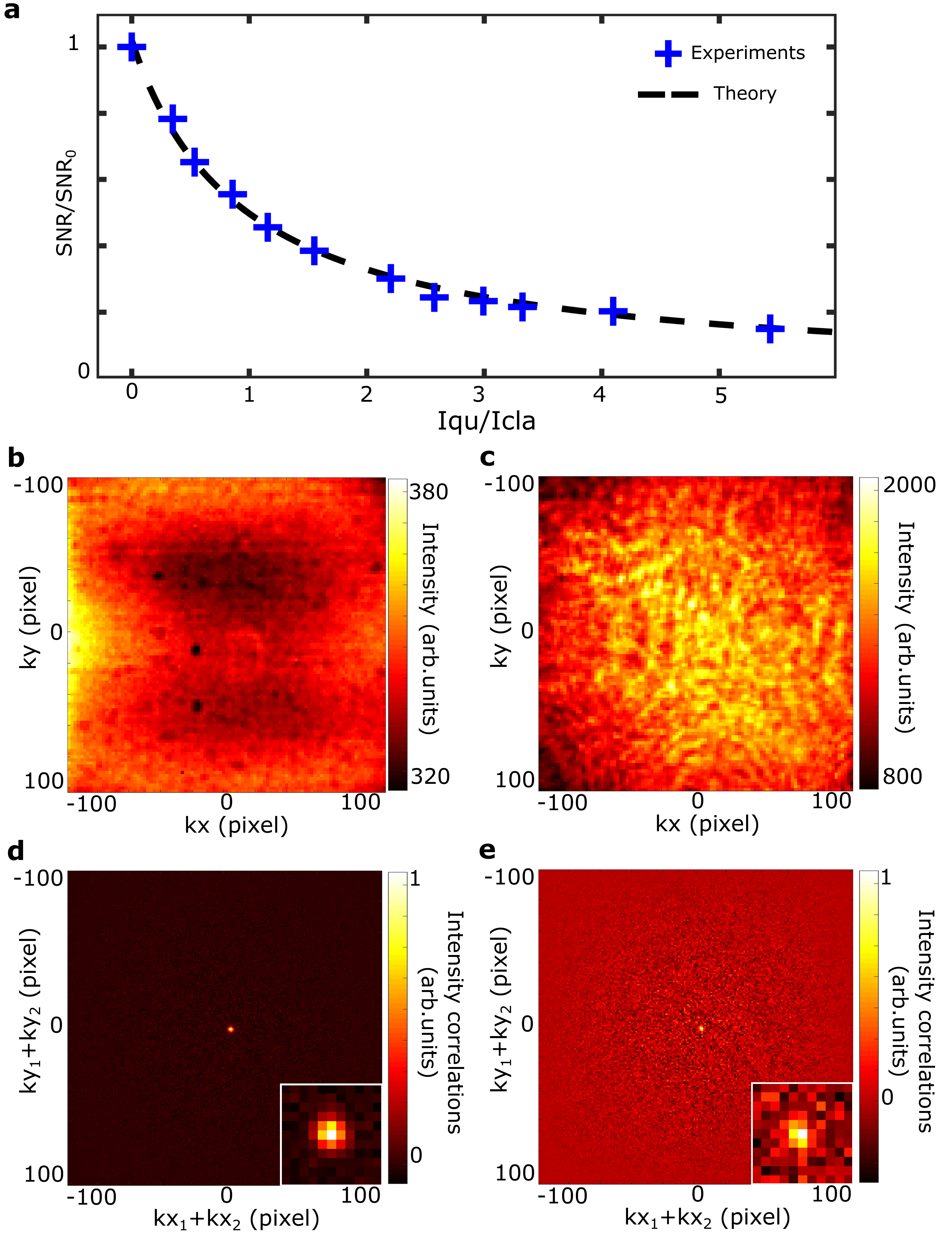} 
\caption{\label{FigureSM10} \textbf{SNR variation in the presence of stray light.} \textbf{a}, Normalised SNR values in function of the average intensity ratio between classical and quantum illumination $\langle I_{cl} \rangle / \langle I_{qu} \rangle$. \textbf{b} and \textbf{c}, Intensity images measured for $\langle I_{cl} \rangle / \langle I_{qu} \rangle = 0$ and $\langle I_{cl} \rangle / \langle I_{qu} \rangle = 5.4$, respectively. \textbf{d} and \textbf{e}, Images of intensity correlation measurements represented in the sum-coordinate axis $\vec{k_1}+\vec{k_2}$ for $\langle I_{cl} \rangle / \langle I_{qu} \rangle = 0$ and $\langle I_{cl} \rangle / \langle I_{qu} \rangle = 5.4$, respectively. Insets are zooms of the central peak area.}
\end{figure*}

We characterise the SNR variation in function of the average intensity of classical light falling on the sensor using a similar approach reported elsewhere~\cite{defienne_quantum_2019}. We use the experimental setup described in Fig.3 of the manuscript without the random phase disorder and with no phase object programmed on the SLM. Intensity correlation measurements are then performed for different classical stray light average intensity values $\langle I_{cl} \rangle$, with a constant quantum illumination intensity $\langle I_{qu} \rangle$. These intensity correlation measurements are visualized by projecting them onto the sum-coordinate axis~\cite{moreau_realization_2012,tasca_imaging_2012}. For example, Fig.~\ref{FigureSM10}.d and e show images of intensity correlation measurements in the sum-coordinate basis in the presence of classical stray light with an average intensity ratios of $\langle I_{cl} \rangle / \langle I_{qu} \rangle = 0$ and $\langle I_{cl} \rangle / \langle I_{qu} \rangle = 5.4$, respectively. The bright peaks at the center of the images are signatures of the strong momentum anti-correlation between the pairs. The peak value is proportional to the mean value of an intensity correlation image measured in the same conditons (i.e.\textit{signal} of the SNR). The standard deviation of noise in the pixels surrounding the peak is proportional to the standard deviation of the noise in an intensity correlation image acquired in the same conditions (i.e. \textit{noise} of the SNR). The SNR values shown in Fig~\ref{FigureSM10}.a are calculated by dividing peak values by standard deviations for different values of $\langle I_{cl} \rangle / \langle I_{qu} \rangle$. Fitting these experimental results with a theoretical model described in~\cite{reichert_optimizing_2018} shows that the SNR varies as  $ 1 / \langle I_{cl} \rangle$.

\section{Details on the resolution enhancement characterisation}

This section provides more details about the comparison of spatial resolution between classical and quantum holographic systems. 

\begin{figure*}
\centering
\includegraphics[width=0.80 \textwidth]{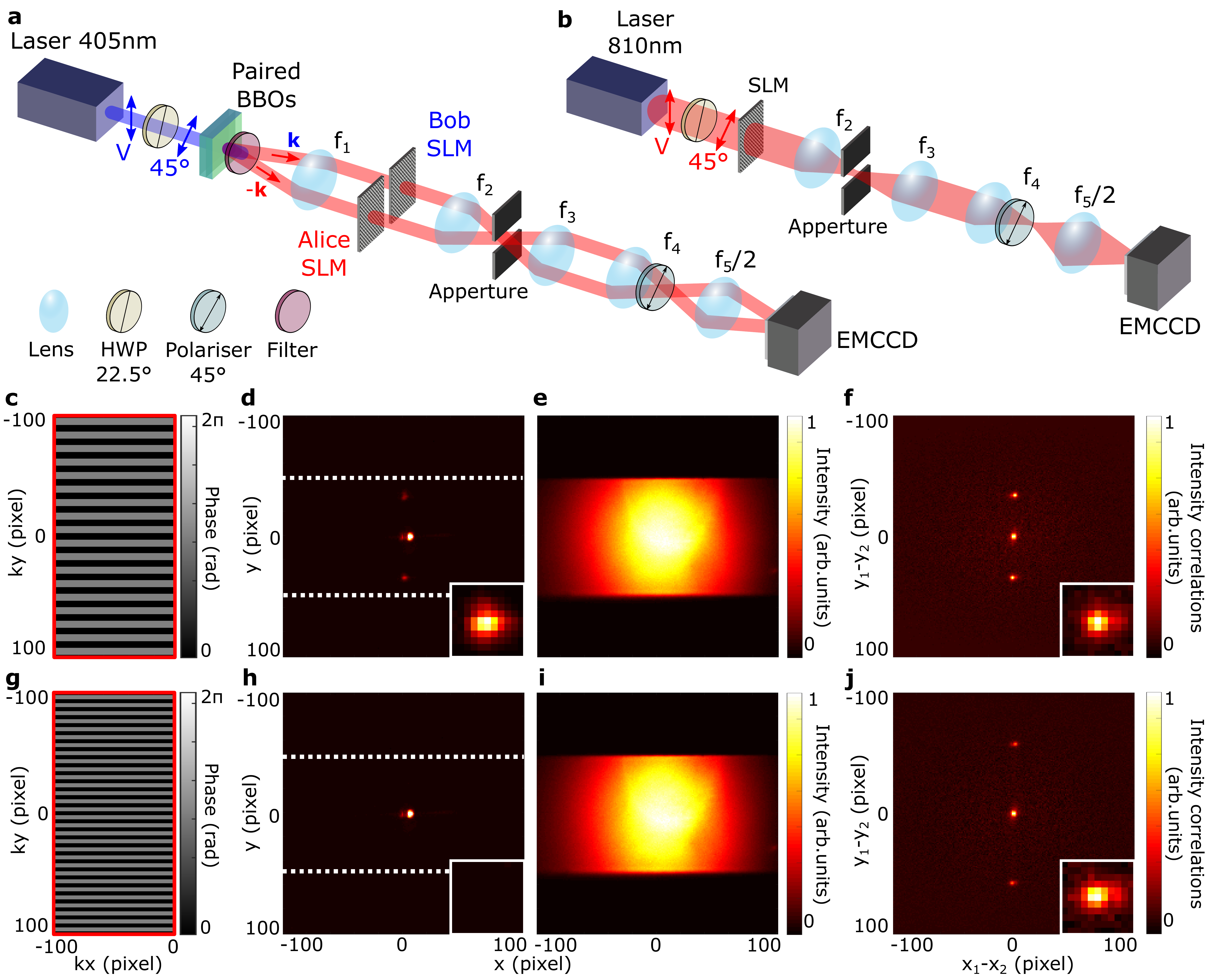} 
\caption{\label{FigureSM8} \textbf{Resolution enhancement characterisation.} \textbf{a} and \textbf{b}, Experimental apparatus used for spatial frequency cut-off measurement of the quantum and classical holographic systems, respectively. \textbf{c}, $26$ pixels period phase grating programmed on the SLM (only Alice SLM in the quantum case). \textbf{d}, Intensity image measured with the classical system. Inset is a zoom on the first-order diffraction peak. White dashed lines represent the edges of the aperture. \textbf{e}, Intensity image measured in the quantum system. \textbf{f}, Projection of the intensity correlation matrix onto the minus-coordinate axis $\vec{r_1}-\vec{r_2}$  that shows three diffraction peaks. Inset is a zoom on the first-order diffraction peak. \textbf{g}, $16$ pixels period phase grating programmed on the SLM. \textbf{h}, Intensity image measured with the classical system. First-order diffraction peaks are blocked by the aperture. \textbf{i}, Intensity image measured in the quantum system. \textbf{j}, Projection of the intensity correlation matrix onto $\vec{r_1}-\vec{r_2}$ that still shows three diffraction peaks.}
\end{figure*}

Various criteria can be used to compare spatial resolution of two imaging systems~\cite{goodman_introduction_2005}. In our work, we chose to compare their spatial frequency cut-off. Figure~\ref{FigureSM8} shows the experimental apparatus used to perform such analysis with the quantum holographic protocol (Figure~\ref{FigureSM8}.a) and its classical version (Figure~\ref{FigureSM8}.b). In these configurations, the Fourier plane of the SLM, that is also the plane of the aperture, is imaged onto the EMCCD camera. This is done by substituting the lens $f_5$ in the initial experimental setup described in Figure 3 of the manuscript by a lens of half focal length $f_5/2$. In the classical case, when a $26$ pixel period phase grating (Fig.~\ref{FigureSM8}.c) is programmed on the SLM (i.e. on Alice SLM in the quantum case), the intensity image shows a diffraction pattern with three main components: a central zero-order peak and two symmetrically positioned plus-or-minus first-order peaks (Fig.~\ref{FigureSM8}.d). In the quantum case, intensity image does not reveal any specific diffraction pattern (Fig.~\ref{FigureSM8}.e). To reveal the frequency content of the phase object under the quantum illumination, we perform intensity correlation measurements between all pair of pixels of the EMCCD camera and visualise these data by projecting them along the minus-coordinate axis $\vec{r_1}-\vec{r_2}$~\cite{defienne_adaptive_2018-3,devaux_quantum_2019}. Similarly to the classical case, we observe three peaks of intensity correlations in Fig.~\ref{FigureSM8}.e. When the same experiment is reproduced with a $16$ pixels period grating (Fig.~\ref{FigureSM8}.g), the first-order diffraction peak is blocked by the aperture in the classical case (Fig.~\ref{FigureSM8}.h), while it remains present under quantum illumination (Fig.~\ref{FigureSM8}.j). In Fig.5.b of the manuscript, red circles and blue crosses are values of (plus) first-order peak intensities measured for different grating periods in the classical and quantum case, respectively. 

The use of spatially entangled photon pairs to enhance the resolution of an imaging system compared to its classical coherent version was previously demonstrated~\cite{boto_quantum_2000}. This effect can be understood using a simple theoretical model~\cite{abouraddy_entangled-photon_2002}. First, we assume that the crystal is very thin and the pump beam diameter is infinite. In this case, the spatial component of the two-photon wavefunction can be written as:
\begin{equation}
\phi(\vec{r_1},\vec{r_2}) = \delta(\vec{r_1}-\vec{r_2})
\end{equation}
where $\vec{r_1}$ and $\vec{r_2}$ are positions in the crystal plane. Then, we consider an object $t(\vec{k})$ positioned into one half of the Fourier plane of the crystal (i.e. illuminated by only one photon of the pair). The object is imaged onto another imaging plane through an imaging system with a coherent point spread function denoted $h(\vec{k})$. The intensity correlation image $R(\vec{k})$ obtained at the output can be written as: 
\begin{equation}
R(\vec{k}) = \abs{t \ast h^2}^2(\vec{k})
\end{equation}
where $\ast$ is the convolution product. In the classical case, the intensity image $I(\vec{k})$ obtained at the output is: 
\begin{equation}
I(\vec{k}) = \abs{t \ast h}^2(\vec{k})
\end{equation}
Because the convolution kernel in the quantum case is squared compared to classical coherent light, the spatial resolution is enhanced by a factor $2$. In the Fourier domain, this resolution enhancement corresponds a broadening of the optical transfer function and therefore an increase of the corresponding spatial frequency cut-off~\cite{goodman_introduction_2005}. Note that, because the crystal has in practice a finite thickness and the pump beam diameter is not infinite, the spatial resolution enhancement measured is lower than $2$ (but higher than $1$).

\section{Details on the photon pair source, detection efficiency of the experimental setup and Allan variance plot}

This section provides more details about the rate of generation of SPDC pairs, the detection efficiency of the setup and reports an Allan variance plot.

\noindent \textbf{Rate of pair generation.} To characterise the photon pair rate of the source, a threshold is applied on the EMCCD camera: in each acquired images, all grey values bellow $210$ are set to $0$ and all values above are set to $1$. As detailed in~\cite{reichert_optimizing_2018}, this approach turns the EMCCD into a multi-pixel photon counter with an effective quantum efficiency of $0.61$. By acquiring a total of $7000$ frames with an exposure time of $3$ms (no pattern on the SLM and after removing the polarisers), we measure an average photon rate of $304$ photons per second per pixel and an average number of pairs of $0.1$ pairs per pixel per second (spatial mode = 1.1 pixels, as shown in~Fig.\ref{FigureSM6}). The same experiment reproduced with the shutter closed returns $4.1$ photons per pixel per second and $5.10^{-4}$ pairs per spatial mode per second, showing that the noise is negligible. 

\noindent \textbf{Detection efficiency of the experimental setup}. Detection efficiency of the entire setup is estimated by multiplying the efficiency of each optical components. The SLM is an Holoeye Pluto NIR-II with a reflection efficiency of approximately $0.64$ at $800$nm. Lenses are all coated for $800$nm, therefore their losses are negligible. Finally, the Andor Ixon Ultra EMCCD camera sensor has approximately $0.75$ quantum efficiency (data sheet provided by Andor). The total detection efficient is therefore about $0.75*0.64 = 0.48$.

\begin{figure*}
\centering
\includegraphics[width=0.9 \textwidth]{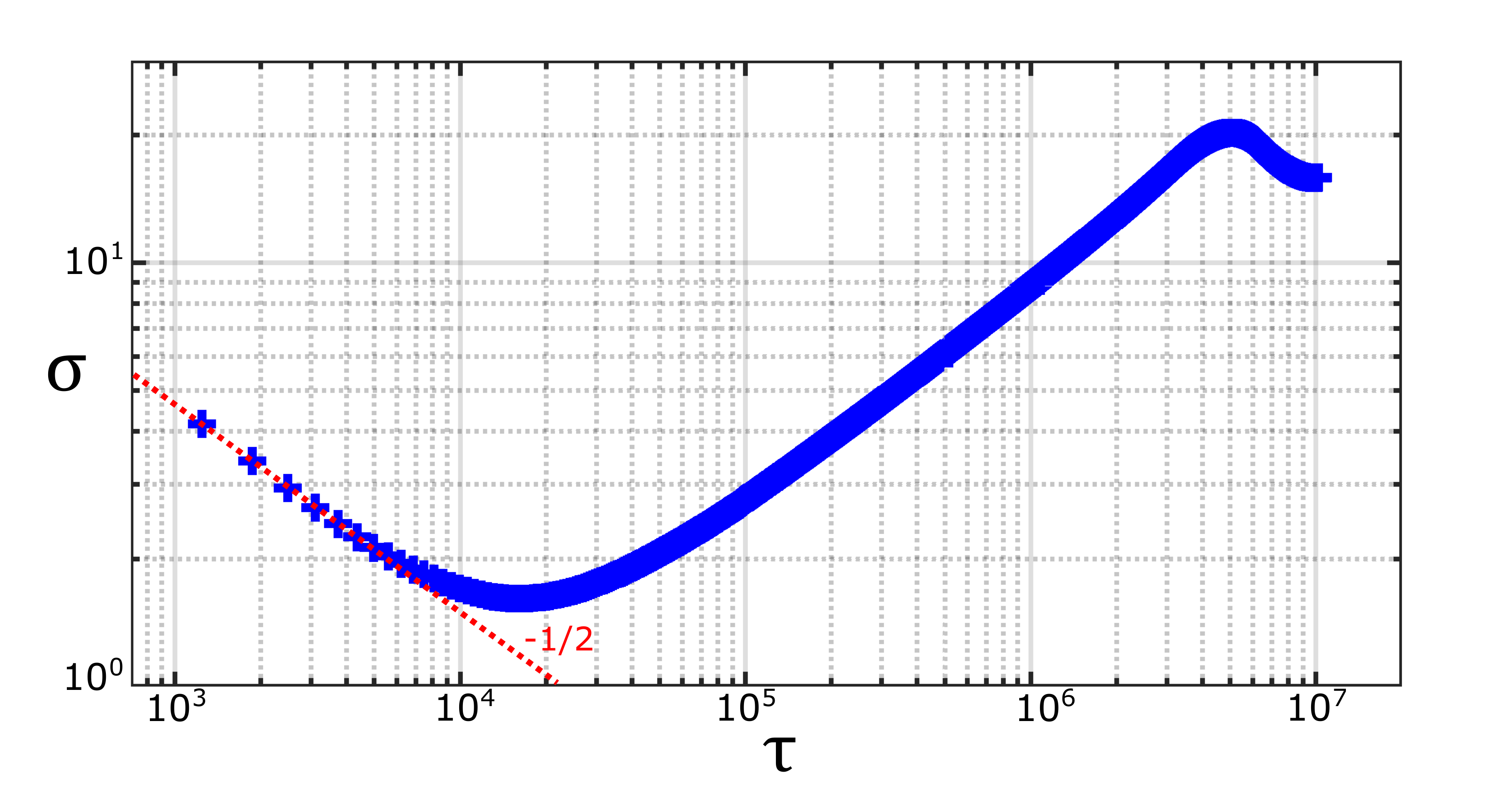} 
\caption{\label{FigureSM12} \textbf{Allan deviation plot}.  Allan deviation values $\sigma$ (blue crosses) are represented in function of the averaging time $\tau$, together with a $-1/2$ slope (red). Time unit is in number of frames.}
\end{figure*}

\noindent \textbf{Allan deviation.} Figure~\ref{FigureSM12} shows an Allan deviation plot~\cite{allan_statistics_1966}. It is constructed from the successive measurement of $N=50000$ intensity correlation images (flat SLM pattern), each measured from $622$ frames, that is the size of the internal buffer of the camera. In our experiment, the 'time' domain signal denoted $R(t)$ is composed of the spatial averaged values (over an area of $100 \times 100$ pixels) of each intensity correlation image taken at an acquisition number $t \in [1,50000]$ (i.e. the 'time' corresponds to the 'image set number'). To calculate the Allan deviation $\sigma(\tau)$ from $R(t)$, where $\tau$ is the averaging time, we used the method detailed in~\cite{variance_noise_2015}. Figure~\ref{FigureSM12} shows that the slope at small $\tau$ equals $-1/2$, meaning that our measurements are not shot-noise-limited (i.e.slope $-1$).

\section{Quantum holographic imaging of real objects}

\begin{figure*}
\includegraphics[width=1 \textwidth]{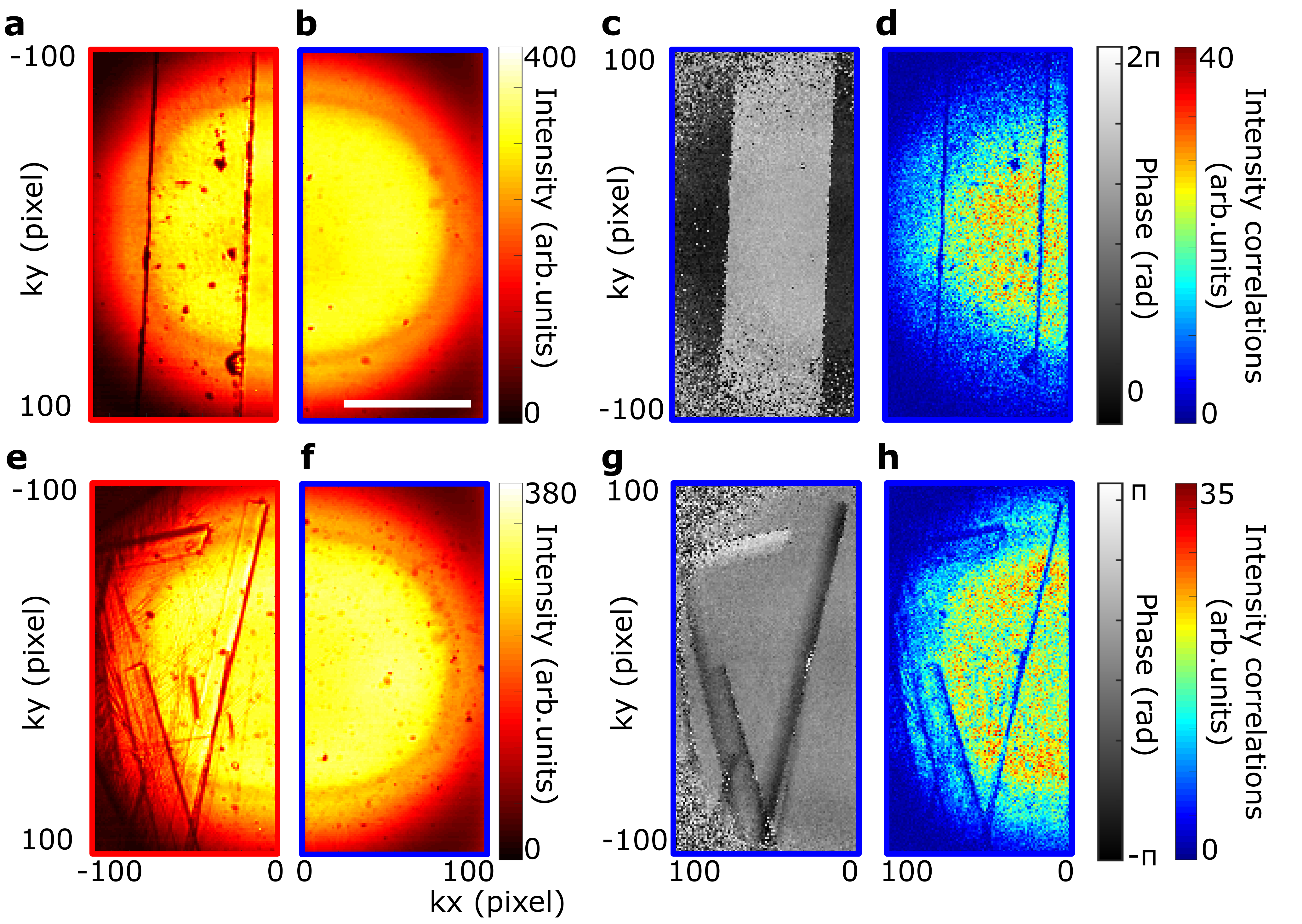} 
\caption{\label{FigureSM15} \textbf{Quantum holographic imaging of real objects.} \textbf{a}, Intensity images measured by Alice showing a piece of transparent scotch tape. \textbf{b}, Intensity image measured by Bob. \textbf{c}, Phase image reconstructed by Bob with SNR$=14$. \textbf{d}, Amplitude image reconstructed by Bob from the same set of intensity correlation images by replacing the argument in equation (1) of the manuscript with an absolute value. \textbf{e}, Intensity image measured by Alice showing parts of a bird feather. \textbf{f}, Intensity image measured by Bob. \textbf{g}, Phase image reconstructed by Bob with SNR$=13$. \textbf{h}, Amplitude image reconstructed by Bob. $10^7$ frames were acquired in total for each case. The white scale bar corresponds to $1$mm. Phase and amplitude images retrieved by Bob are rotated by 180 degrees for convenience.}
\end{figure*}

Figure~\ref{FigureSM15} shows results of quantum holographic imaging performed on two real objects, a piece of scotch tape and a bird feather. The objects are positioned on a microscope slide in place of Alice SLM. Spatial phase variations due to the stress-induced (scotch tape) and structural (bird feather) birefringence are clearly visible in the phase images reconstructed by Bob in Figs.~\ref{FigureSM15}.c and g. Furthermore, we show that Bob can also retrieve the object amplitudes (Figs.~\ref{FigureSM15}.d and h) from the same set of intensity correlation measurements by simply replacing the argument in equation (1) of the main manuscript with an absolute value. These results show that our quantum holographic approach can be used as a practical tool for phase and amplitude measurements of complex objects, including biological samples.

\begin{figure*}
\centering
\includegraphics[width=0.6 \textwidth]{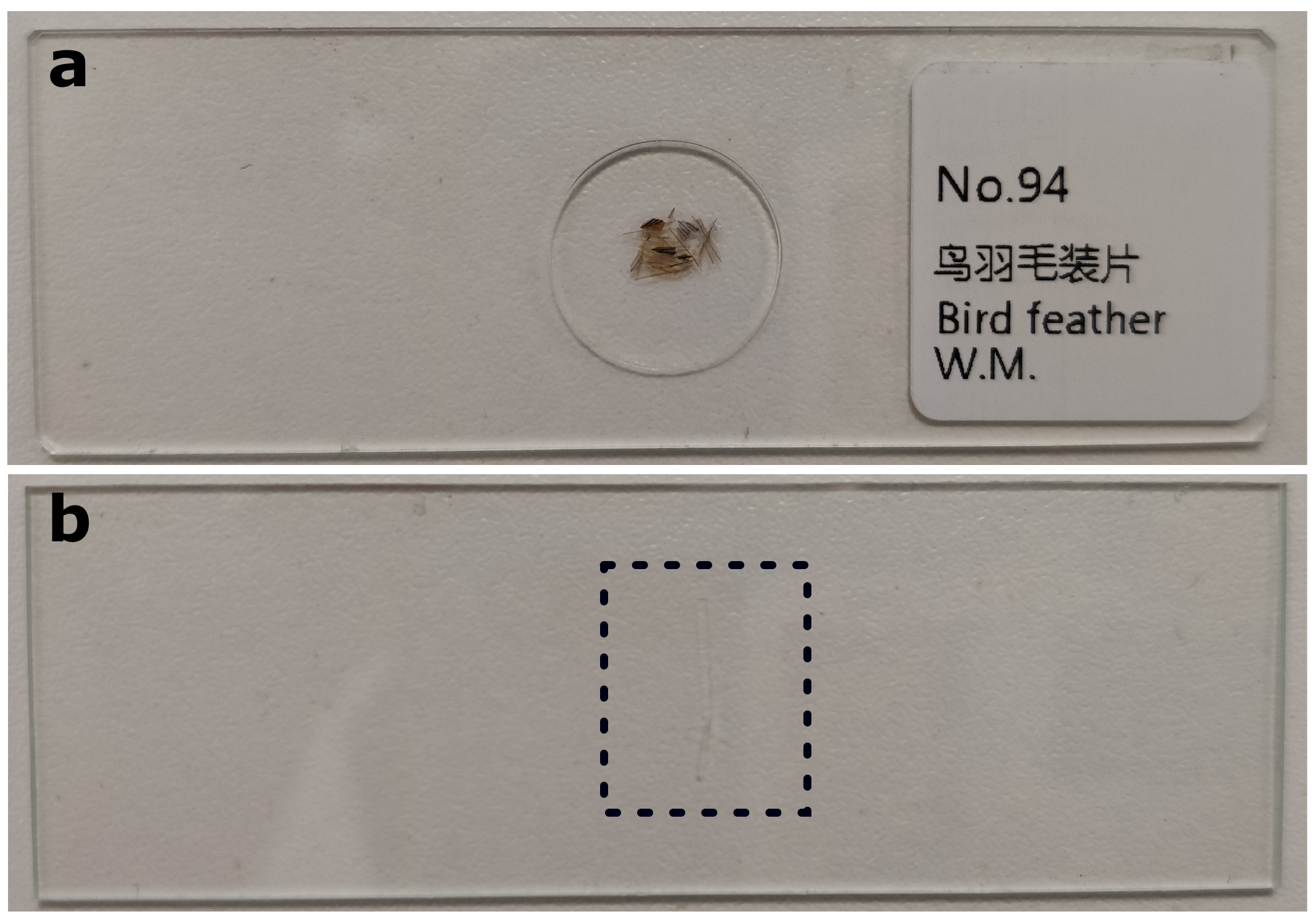} 
\caption{\label{FigureSM14} \textbf{Photos of microscope slides}. \textbf{a}, Bird feather on a microscope slide provided by the company KKmoon. \textbf{b}, Piece of scotch tape on a microscope slide from the company Sellotape.}
\end{figure*}

Figure~\ref{FigureSM14} shows photos of the microscope slides containing the bird feather and the piece of scotch tape.

Our quantum holographic scheme is currently limited to image the phase of polarisation-sensitive objects. To extend its range of application, Figure~\ref{FigureSM13}.a shows a modified version of our experimental setup that can be used to achieve quantum holography with non-polarisation sensitive object. This configuration harnesses concepts from Differential Interference Contrast microscopy~\cite{allen_zeiss-nomarski_1969}. In this new configuration, Alice SLM is not used anymore (i.e. flat phase mask shown in Figure~\ref{FigureSM13}.b) and is replaced by a non-birefringent sample located in a conjugated image plane positioned between lenses $f_3$ and $f_4$. Furthermore, two Savart plates are positioned just before and after the sample: the first plate separates the incident beam into two parallel beams and the second plate recombine them. In our experiments, the Savart plates are made of two layers of quartz of 3 mm thickness, with optical axes tilted at 45 degrees with respect to the surface normal. Each SP produces a shear of approximately $25$ $\upmu$m. As detailed in~\cite{terborg_ultrasensitive_2016}, such arrangement produces a polarisation-sensitive phase pattern at its output that is directly proportional to the spatial phase of the non-birefringent sample inserted between the plates. Figures~\ref{FigureSM13}.c-h show preliminary results of phase measurement of a random phase layer fabricated by spraying some silicone adhesive onto a microscopic slide. When using our conventional experimental (without the Savart plates), we clearly observe that the sample is not polarisation dependent because no phase variation are visible in the reconstructed phase in Figure~\ref{FigureSM13}.e. In contrary, the phase measured using the new experimental configuration in Figure~\ref{FigureSM13}.g shows random spatial variations of the phase, confirming that our quantum holographic scheme is now sensitive to non-polarisation sensitive objects. 

\section{Phase imaging through static phase disorder} 
\begin{figure*}
\centering
\includegraphics[width=1 \textwidth]{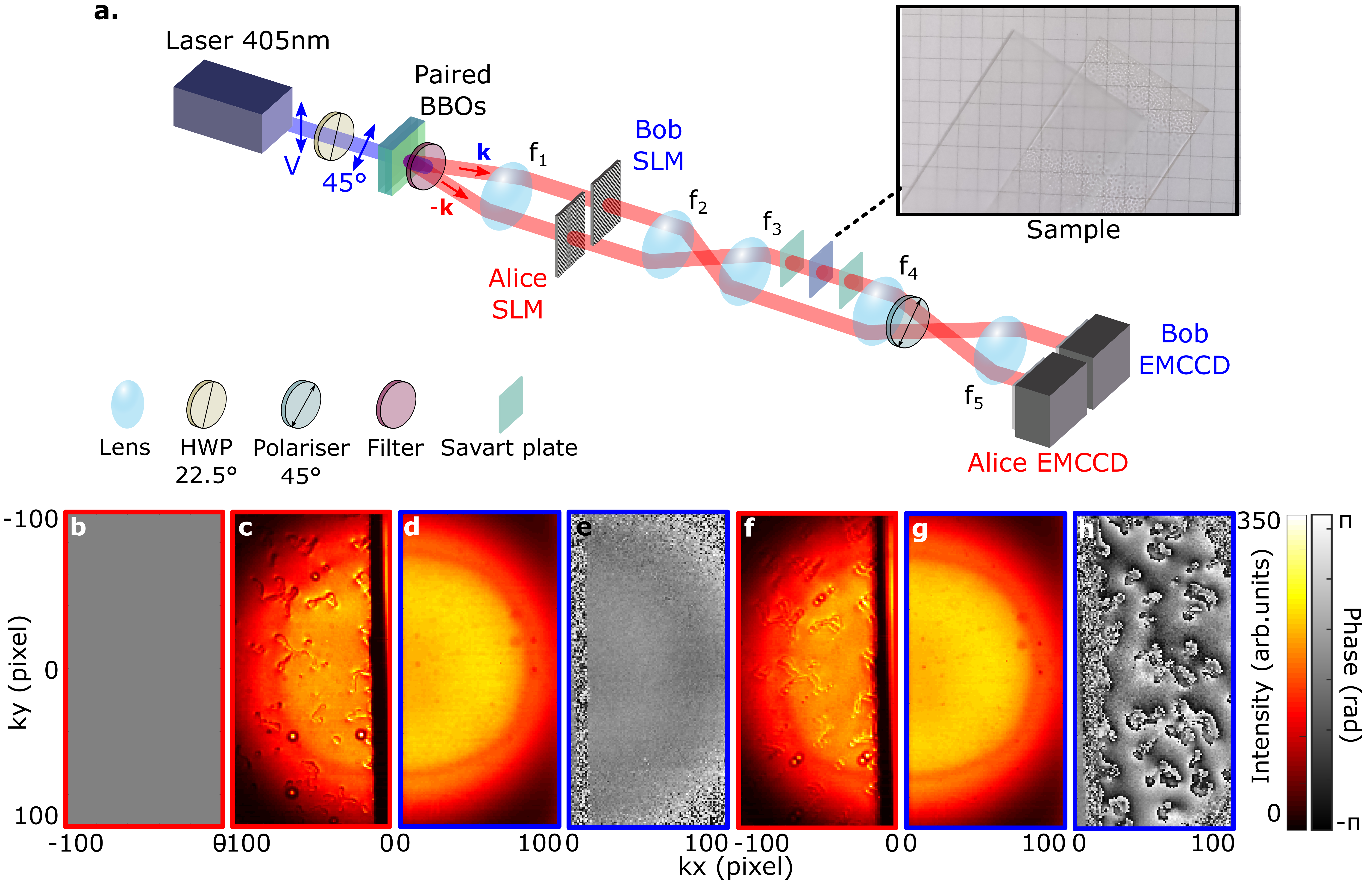} 
\caption{\label{FigureSM13} \textbf{Quantum holography of non-polarisation sensitive objects}. \textbf{a}, Modified quantum holographic setup to achieve phase imaging of non-polarisation sensitive phase object. The sample is inserted in a conjugate image plane of Alice SLM located between $f_3$ and $f_4$. Two Savart plates are inserted on each side of the sample and are slightly tilted. The sample is a microscopic slide covered by a layer of silicone adhesive generated using a spay, that effectively produces a random phase layer.  \textbf{b}, Flat phase pattern on Alice SLM (not in use in this configuration), \textbf{c} and \textbf{d}, Intensity images measured by Alice and Bob without the Savart plates. \textbf{e}, Phase reconstructed by Bob using the quantum holographic approach without the Savart plates with SNR$=17$. \textbf{f} and \textbf{g}, Intensity images measured by Alice and Bob with the Savart plates. \textbf{h}, Phase reconstructed by Bob using the quantum holographic approach with the Savart plates with SNR$=12$. Each phase image was reconstructed from $5.10^6$ frames.}
\end{figure*}

Figure~\ref{FigureSM13}.e shows a phase image reconstructed using the conventional quantum imaging setup (i.e. setup in Figure~\ref{FigureSM13}.a without the Savart plates) of a flat phase object programmed on Alice SLM (Figure~\ref{FigureSM13}.b) in the presence of a static random non-birefringent phase disorder on the optical path. The absence of any phase variations in Figure~\ref{FigureSM13}.e shows that our holographic protocol is also insensitive to the presence of static phase disorder, thus confirming that the dynamics of the disorder has no influence (e.g. as an averaging effect) on the robustness of the phase disorder advantage demonstrated in Figure 3 in the manuscript.

\section{Details on decoherence free subspace and collective dephasing} 

By definition, a decoherence free subspace (DFS) is a subspace of the Hilbert space of a quantum system that is invariant to non-unitary dynamics e.g. dynamic dephasing. In our experiment, the subspace spanned by the states $\ket{HH}$ and $\ket{VV}$ is invariant to a dynamic dephasing process, that is a non-unitary evolution operator, and is then by definition a DFS~\cite{lidar_decoherence-free_2003}. 

The notion of DFS is strongly linked to the properties of the decoherence effect. The decoherence effect considered in our work is a type of \textit{collective} dephasing process. A general dephasing process corresponds to a qubit state $\ket{P}_k$ (where $P$ and $k$ are the polarisation and spatial mode of a photon, respectively) that undergoes a random phase shift $\phi_{P,k}$ varying over time: $\ket{P}_k \rightarrow e^{i \phi_{P,k}} \ket{P}_k$. Such process is called \textit{collective} if they are some correlations between the random phases applied to the different qubits that persist over time. For example, in many previous works on DFS~\cite{kwiat_experimental_2001,banaszek_experimental_2004}, these correlations are spatial: two different qubits $\{ \ket{P_1}_{k_1},\ket{P_2}_{k_2} \}$ undergo random phase shifts $\phi_{P_1,k_1}$ and $\phi_{P_2,k_2}$ that vary over time but remain correlated in space at any time i.e. $\phi_{P,k_1} = \phi_{P,k_2}$, where $P$ denotes indifferently $P_1$ or $P_2$. While this form of decoherence is relevant in quantum information processing and communication applications, it is less realistic in our imaging configuration. In our case, the presence of two non-correlated non-birefringent dynamic random phase disorder on two different optical paths produces another form of collective decoherence in which the random phases are not correlated in space but in polarisation i.e. $\phi_{P_1,k} = \phi_{P_2,k}$, where $k$ can be indifferently $k_1$ or $k_2$.

\section{Details on imaging in the presence of classical stray light using quantum illumination (QI)}

The QI protocol was introduced by S.Lloyd~\cite{lloyd_enhanced_2008} and extended to Gaussian states by Tan et al~\cite{tan_quantum_2008}, in which a practical version of the protocol is proposed. In imaging, the protocol has been suggested for detecting the presence of an object embedded within a noisy background, even in the presence of environmental perturbations and losses~\cite{guha_gaussian-state_2009,pirandola_advances_2018}. In 2013, Lopaeva et al.~\cite{lopaeva_experimental_2013} performed the first experimental demonstration of the QI protocol to determine the presence an object by exploiting intensity correlations between photons produced by parametric down conversion. Very recently, spatial correlations between downconverted photon pairs have been used to implement full-field QI protocols ~\cite{defienne_quantum_2019,gregory_imaging_2020}, where amplitude objects were reconstructed through coincidence measurements performed by a camera in the presence of static stray light. In our work, the QI protocol is based on similar arrangements than those used in~\cite{gregory_imaging_2020} i.e. one photon of an entangled pair illuminates an object while the other is used as an 'ancilla', and both photons are detected in coincidence using a camera in the presence of classical light. However, our QI protocol goes beyond these previous works because it operates for phase objects (that are completely invisible on the intensity images, as shown for example in Figure I where only the cat-shaped objects are visible) and also works in the presence of dynamic stray light i.e. a time-varying complex speckle pattern superimposed on the sensor. Note also that another difference is that our experiment employs a source of photon pairs entangled not only in space but also polarisation, in which the spatial entanglement is certainly at the base of the QI advantage, but where polarisation entanglement is still required for the holographic process.
 
 \newpage

\bibliography{Biblio}

\end{document}